\documentclass[11pt,a4paper]{amsart}

\usepackage{mathptmx} % Postscript fonts

\usepackage{amsmath,amssymb,latexsym,amsthm,enumerate}
\usepackage{graphicx} %\includegraphics for .ps, .eps (dvips); .pdf, .jpg, .png (pdflatex)
\usepackage{hyperref} %To be able to click on the links in the paper
\usepackage{xcolor} %Colors in text

\usepackage{verbatim} % commands like \begin{comment} \end{comment} do not work without it

\usepackage[font=footnotesize,justification=raggedright,parskip=5pt]{caption,subfig}

\def\NN{\mathbb{N}}
\def\ZZ{\mathbb{Z}} 

\def\RR{\mathbb{R}}
\def\CC{\mathbb{C}}
\def\PP{\mathbb{P}}
\def\II{\mathbb{I}}
\def\EE{\mathbb{E}}

\def\L{\mathcal{L}}

\def\G{{\mathcal G}}

\def\cu{\mathtt{i}}
\def\i{{\sqrt{-1}}}
\def\I{\mathbf{1}}

\newcommand{\olX}{\overline X}  

\newcommand{\olx}{\overline x}  
\newcommand{\oly}{\overline y}
\newcommand{\olS}{\overline S}

\newtheorem{thm}{Theorem}[section]

\newenvironment{defin}{\vspace{1mm}\noindent\textbf{Definition.}}{\vspace{1mm}}

\newenvironment{pr}{\vspace{1mm}\noindent\textbf{Proof.}}
                   {\vspace{-5mm}\begin{flushright}$\Box$\end{flushright}}

\begin{document}

\title{Volatility derivatives in market models with jumps}

\author{Harry Lo}
\address{Imperial College London and Swiss Re}
\email{lo.harry@gmail.com}
\author{Aleksandar Mijatovi\'{c}}
\address{Department of Mathematics, Imperial College London}
\email{a.mijatovic@imperial.ac.uk}

\begin{abstract}
It is well documented that 
a model for the underlying asset price process 
that seeks to capture the behaviour
of the market prices of vanilla options
needs to exhibit
%asset returns can exhibit large fluctuations over very short time intervals.
%It is also generally accepted that 
both diffusion and jump features.
In this paper we assume that the asset price process 
$S$
is Markov
with 
c\`adl\`ag
paths
and propose a scheme for computing
the law of the realized variance of the log returns
accrued 
while the asset was trading in a prespecified corridor.
%quadratic variance of the underlying process,
We thus obtain an algorithm for pricing and hedging 
volatility derivatives and derivatives on the 
corridor-realized variance
in such a market.
The class of models under consideration is large, as it
encompasses
jump-diffusion and 
L\'evy processes.
%the stochastic volatility models with jumps
%described in~\cite{CarrMadanGemanYor_LocalLevy}.
We prove
the weak convergence of the scheme 
and describe in detail
the implementation of the algorithm 
in the characteristic cases
where
$S$
is
a CEV process
(continuous trajectories),
a variance gamma process
(jumps with independent increments) 
or an
infinite activity jump-diffusion
(discontinuous trajectories with dependent increments).
\end{abstract}

\thanks{We would like to thank Martijn Pistorius for many useful discussions.}

\maketitle

\section{Introduction}
\label{sec:Intro}
Derivative securities on the realized variance 
of the log returns of an underlying asset price process
trade actively  in the financial markets.
Such derivatives play an important role 
in risk management  and are also used 
for expressing a view on the future behaviour
of volatility in the underlying market. 
Since the 
liquid contracts
have both 
linear (variance swaps) and
non-linear
(square-root = volatility swaps,
hockey stick = variance options)
payoffs,
it is very important to have a robust algorithm 
for computing 
the entire law of the realized variance. 
Often such contingent claims have an additional feature, 
which makes
them cheaper and hence more attractive to the investor,
that stipulates that variance of log returns accrues 
only when the spot price is trading in a contract-defined
corridor (see Subsection~\ref{subsec:contracts} for the precise
definitions of such derivatives).

It is clear from these definitions that, in order to manage the
risks that arise in the context of volatility derivatives, 
one needs to apply the same modelling framework that
is being used for pricing and hedging vanilla options 
on the underlying asset. It has therefore been argued that the pricing and hedging
of volatility derivatives should be done using models with
jumps %(see~\cite{CarrMadanGemanYor_Variance})
and stochastic volatility (see for example~\cite{Gatheral}, Chapter~11).
In this paper we propose a scheme for computing the distribution
of the realized variance and the corridor-realized variance when
the underlying process 
$S=(S)_{t\geq0}$
is a Markov process with possibly discontinuous trajectories, 
thus obtaining 
an algorithm  for pricing and hedging all the payoffs mentioned 
above.
Our main assumption
%, which is implicit in the
%preceding sentence, 
is that the Markov dimension of 
$S$
is equal to one
%like in~\cite{CarrMadanGemanYor_Variance},
(i.e. we assume
that the future and past of the process
$S$
are independent given its present value).
We do not make any additional assumptions on the structure of the
increments or the distributional properties of the process
$S$.
This class of processes is large as it encompasses one dimensional jump-diffusions
and L\'evy processes. 
%(e.g. the widely used models in~\cite{CarrMadanGemanYor_CGMY}).
%and~\cite{MCC}).

The algorithm consists of two steps. 
In the first step
%the dynamics of 
the original Markov process 
$S$
under a risk-neutral measure
is approximated by a continuous-time finite state
Markov chain
$X=(X_t)_{t\geq0}$.
This is achieved by approximating the 
generator of 
$S$
by a generator matrix for 
$X$.
The second step consists of pricing the corresponding volatility derivative
in the approximate model 
$X$.
It should be stressed that the two steps are independent of each other
but both clearly contribute to the accuracy of the scheme.
In other words the second step can be carried out for any approximate generator
matrix of the chain
$X$.
In specific examples in this paper we describe a natural way
of defining 
the approximate generator matrix (see Section~\ref{sec:second_moments}
for diffusions and Section~\ref{sec:third_moments}
for processes with jumps) 
which is by no means optimal
%but do not investigate in detail how to optimize this choice
(see monograph~\cite{Kurtz}
for weak convergence of such approximations and~\cite{Mijatovic}
for possible improvements)
but already makes the proposed scheme accurate enough
(see the numerical results in Section~\ref{sec:numerical}).

In the second step of the algorithm 
we approximate the dynamics of the corridor-realized
variance of the logarithm of the chain 
$X$
(i.e. the variance that accrued while 
$X$
was in the prespecified corridor)
by a Poisson process
with 
an intensity that is a function of the current 
state of the chain 
$X$.
This approximation is obtained by  
matching 
$k\in\NN$
instantaneous conditional moments
of the corridor-realized variance of the chain 
$X$.
This is a generalisation 
of the method proposed in~\cite{ALM},
which in the framework of this paper
corresponds to 
$k=1$
and
only works in the case of
linear payoffs on
the realized variance
(i.e. variance swaps)
as can be
seen in Tables~\ref{t:cev_varpayoff},
\ref{t:vg_varpayoff} and~\ref{t:SubCEV_varpayoff} 
of Section~\ref{sec:numerical}.
Using
$k$
strictly larger than one improves considerably the
quality of the approximation to the distribution
of the corridor-realized variance for 
$S$.
In fact
if 
$S$
is a diffusion process, then our algorithm with
$k=2$
produces prices for the non-linear volatility payoffs
(e.g. volatility swaps and options on variance)
which are within a few basis points of the true price
(see Table~\ref{t:cev_varpayoff}
and Figure~\ref{fig:cev_var_pdf}).
If the trajectories of 
$S$
are discontinuous, 
then 
the scheme with
$k=3$
appears to suffice
(see Tables~\ref{t:vg_varpayoff} and~\ref{t:SubCEV_varpayoff}
and Figures~\ref{fig:vg_var_pdf} and~\ref{fig:cevvg_var_pdf}).
Note also that in~\cite{LM_Code}
we provide a straightforward implementation of our
algorithm in Matlab 
for
$k=3$.
Furthermore
in Section~\ref{sec:Conv}
we prove 
the weak convergence of 
our scheme 
as
$k$
tends to infinity
(see Theorem~\ref{thm:Conv}).

The general approach of this paper is to view continuous-time
Markov chains as a numerical tool that is based on probabilistic principles
and can therefore be applied in a very natural way to problems in pricing
theory. It is worth noting that 
there is no theoretical obstruction 
for extending our scheme
%, which was
%briefly described in the
%previous two paragraphs, 
to the case when 
$S$
is just one component of a 
two dimensional 
Markov process
(e.g. 
$S$
is the asset price in a
stochastic volatility model)
by 
%approximating this two dimensional
using a Markov chain to approximate this two dimensional
process.
The reason why 
throughout this paper
we assume that 
$S$
itself
is Markov 
lies in the 
feasibility of the associated  
numerical scheme.
If
$S$
is Markov
the dimension of the generator of 
$X$
can be as small as 
$70$,
while in the case of the stochastic volatility process
we would need to find the spectra of matrices of dimension
larger than 
$2000$.
This is by no means impossible but is not the focus of the present paper. 

The literature on the pricing and hedging of derivatives 
on the realized 
%and corridor-realized 
variance is vast. 
It is generally agreed that either the assumption
on the independence of increments or the continuity of trajectories of the
underlying process needs to be relaxed in order to obtain a realistic
model for the realized variance. 
In the recent paper~\cite{CarrLee}
model independent bounds for options on variance are obtained
in a general continuous semimartingale market.
The continuity assumption is relaxed in~\cite{CarrMadanGemanYor_Variance},
where a class of one dimensional Markov processes with
independent increments
is considered and the law of the realized variance is obtained.
A perfect replication  for a corridor variance swap 
(i.e. the mean of the corridor-realized variance)
in the case  of
a continuous asset price process is given in~\cite{CarrMadan_Towards}.
For other contributions to the theory of volatility derivatives
see~\cite{ALM} and the references therein.
The main aim of this paper is to provide a stochastic approximation
scheme for the pricing and hedging of derivatives on the realized (and corridor-realized) 
variance in models that violate both the above assumptions,
thus making it virtually impossible to find the laws of the relevant random variables in 
semi-analytic form.

The paper is organised as follows. Section~\ref{sec:k_moments}
defines the approximating Markov chains and gives a general description of 
the pricing algorithm. In Section~\ref{sec:Conv} we state and prove the weak convergence
of the proposed scheme. Section~\ref{sec:third_moments} (resp.~\ref{sec:second_moments})
describes the implementation of the algorithm 
in the case where the process
$S$
is an infinite activity jump-diffusion 
(resp. 
has continuous trajectories).
Section~\ref{sec:numerical} 
contains numerical results and Section~\ref{sec:conclusion}
concludes the paper.

\section{The $k$ conditional moments of the realized variance}
\label{sec:k_moments}
Let
$S=(S_t)_{t\geq0}$
be a strictly positive Markov process with c\`adl\`ag
paths (i.e. each path is right-continuous  as a function of time
and has a left limit at every time
$t$)
which serves as a
model for the evolution
of the risky security under a risk-neutral measure.
Note that
we  are also implicitly assuming that 
$S$
is a semimartingale.

\subsection{The contracts}
\label{subsec:contracts}
A 
\textit{volatility derivative} 
in this market is any security that pays
$\phi([\log(S)]_T)$
at maturity
$T$,
where 
$\phi:\RR_+\to\RR$
is a measurable payoff function and 
$[\log(S)]_T$
is the 
%realised variance at a maturity %$T$
is the quadratic variation 
of the process
$\log(S)=(\log(S_t))_{t\geq0}$
at maturity
$T$
defined by
\begin{eqnarray}
\label{eq:QuadVarS}
[\log(S)]_T & := & \lim_{n\to\infty}\sum_{t_i^n\in\Pi_n, i\geq1} \left(\log\frac{S_{t_i^n}}{S_{t^n_{i-1}}}\right)^2,
\end{eqnarray}
where
$\Pi_n=\{t_0^n, t_1^n,\ldots,t_n^n\}$,
$n\in\NN$,
is a refining sequence of partitions
of the interval
$[0,T]$.
In other words 
$t_0^n=0$,
$t_n^n=T$,
$\Pi_n\subset\Pi_{n+1}$
for
all
$n\in\NN$
and
$\lim_{n\to\infty}\max\{|t_i^n-t_{i-1}^n|:i=1,\ldots,n\}=0$.
It is well-known that this sequence converges in probability
(see~\cite{JacodShiryaev}, Theorem 4.47).
Many such derivative products trade actively in financial markets
across asset-classes (see~\cite{ALM} and the references therein).

A \textit{corridor variance swap}
is a 
derivative security with a linear payoff function that
depends on the accrued variance of the asset price
$S$
while it is trading in an 
interval
$[L,U]$
that
is specified in the contract,
where
%$L,U\in[0,\infty]$
%that satisfy
$0\leq L<U\leq\infty$.
More specifically 
if we define a process
\begin{eqnarray}
\label{eq:MinMaxFun}
\olS_t:=\max\{L,\min\{S_t,U\}\}, \qquad\forall t\in[0,\infty),
\end{eqnarray}
then for
a given partition 
$\Pi_n=\{t_0^n, t_1^n,\ldots,t_n^n\}$
of the time interval
$[0,T]$
%as described in the previous paragraph,
the \textit{corridor-realized variance} is given by
%random variable of interest can be expressed as
\begin{eqnarray}
\label{eq:CorrVarSwap}
%[\log(S)]_T & := & \lim_{n\to\infty}
& & \sum_{t_i^n\in\Pi_n, i\geq1}\left[\I_{[L,U]}(S_{t^n_{i-1}})+\I_{[L,U]}(S_{t^n_{i}})- \I_{[L,U]}(S_{t^n_{i-1}}) \I_{[L,U]}(S_{t^n_{i}})
\right] \left(\log\frac{\olS_{t_i^n}}{\olS_{t^n_{i-1}}}\right)^2, 
\end{eqnarray}
where 
$\I_{[L,U]}$
denotes the indicator function of the interval
$[L,U]$.
In practice the increments
$t^n_{i}-t^n_{i-1}$
ususally 
equal 
one day.
The square bracket in the sum 
in~\eqref{eq:CorrVarSwap}
ensures that the accrued 
variance is not increased when the asset price
$S$
jumps over the interval
$[L,U]$.

The one sided corridor-realized variance was defined in~\cite{CarrLewis}.
%using 
%formulae~(1.1) and~(1.2).
Definition
(1.1)
in~\cite{CarrLewis}
(resp. (1.2)
in~\cite{CarrLewis})
corresponds to 
expression~\eqref{eq:CorrVarSwap}
above
if we choose
$U=\infty$
(resp.
$L=0$).
Formulae~(1.1) and~(1.2)
in~\cite{CarrLewis} 
are used to define 
the corridor-realized variance
in a way which treats the entrance of 
$S$ 
into the corridor 
differently from its exit from the corridor.
This asymmetry is then exploited to 
obtain an approximate hedging strategy for 
linear payoffs on the corridor-realized variance. 
%(i.e. 
%an approximation to the hedge for a corridor variance swaps).
In this paper we opt for a symmetric treatment of the entrance and
exit
of
$S$
into and from the corridor
$[L,U]$,
because this is in some sense more natural. 
It is however important 
to note that all the theorems and the algorithm 
proposed here do NOT depend in any significant way on this choice of definition. 
%in
%expression~\eqref{eq:CorrVarSwap}.
In other words for any reasonable modification of the 
definition in~\eqref{eq:CorrVarSwap}
%corridor-realized variance
(e.g. the one 
in~\cite{CarrLewis})
the algorithm described in this section would still work.
Note also that our algorithm will yield
an approximate distribution
of random variable~\eqref{eq:CorrVarSwap} 
in the model 
$S$
and therefore allows us to 
price any non-linear payoff that depends on the 
corridor-realized variance. 

%It is generally accepted that in the case of 
%volatility derivatives, the quadratic variation defined in~\eqref{eq:QuadVarS}
%is used as a proxy for the discretely observed realized variance.
In the case the corridor-realized variance is monitored continuously
(see~\cite{CarrMadan_Towards}),
we can express it 
using quadratic variation 
as follows. Note first that since the map 
$s\mapsto\max\{L,\min\{s,U\}\}$
can be expressed
as a difference of two convex functions,
Theorem~66 in~\cite{Protter}
implies that the process
$\olS=(\olS_t)_{t\geq0}$
is again a semimartingale
and therefore the corridor-realized variance 
$Q^{L,U}_T(S)$,
defined as the limit 
of the expression in~\eqref{eq:CorrVarSwap}
as
$n$
tends to infinity,
exists and
equals
\begin{eqnarray}
\label{eq:CorrRealVar}
Q^{L,U}_T(S) & = &
[\log(\olS)]_T-\left(\log\frac{U}{L}\right)^2\sum_{0\leq t\leq T}
\left[\I_{(0,L)}(S_{t-}) \I_{(U,\infty)}(S_{t})+
\I_{(0,L)}(S_{t}) \I_{(U,\infty)}(S_{t-})\right]
\end{eqnarray}
by Theorem~4.47a in~\cite{JacodShiryaev}.
Since we are assuming that the process
$S$
is c\`adl\`ag
the limit
$S_{t-}:=\lim_{s\nearrow t}S_s$
exists almost surely for all
$t>0$.
The sum in~\eqref{eq:CorrRealVar},
which
corresponds to jumps of the asset price
$S$
over the corridor
$[L,U]$,
is almost surely finite 
by Theorem~4.47c in~\cite{JacodShiryaev}.
Note also that if
$L=0$
(resp. 
$U=\infty$)
we find that
$Q^{0,U}_T(S)$
(resp.
$Q^{L,\infty}_T(S)$)
equals
the quadratic variation of the semimartingale
$\log(\olS)=(\log(\olS_t))_{t\geq0}$
because
the process
$S$
cannot in these cases jump over the corridor.
Our main task it to find an approximate law of the random 
variable 
$Q^{L,U}_T(S)$
which will allow us to price any derivative  
on the corridor-realized variance
with 
terminal value
$\phi(Q^{L,U}_T(S))$,
where
$\phi$
is a possibly non-linear function.

\subsection{Markov chain \boldmath $X$ and its corridor-realized variance}
%The main assumption 
%in this paper is
Let us start by assuming that 
we are given a generator matrix
$\L$
of a continuous-time Markov chain
$X=(X_t)_{t\geq0}$
which approximates the generator of the Markov process
$S$.
%By slightly abusing the nontation we define 
The state-space
of the Markov chain 
$X$
is
the  set
$E:=\{x_0,\ldots,x_{N-1}\}\subset \RR_+$
with
$N\in\NN$
elements,
such that
$x_i<x_j$
for any integers
$0\leq i<j\leq N-1.$
In Sections~\ref{sec:second_moments}
and~\ref{sec:third_moments}
we discuss briefly how to construct such approximate generators
for Markov processes that are 
widely used in finance  
(i.e. diffusion 
processes with jumps.)
%Our main aim in this paper is to approximate the dynamics of the process
Throughout the paper 
we will 
use the
notation
$\L(x,y)=e_x'\L e_y$
for the elemetns of the matrix
$\L$,
where
$x,y\in E$,
vectors
$e_x,e_y$
denote the corresponding standard basis vectors
of
$\RR^N$
and
$'$
is transposition.

The quantities of interest are 
the quadratic variation 
$[\log(X)]=([\log(X)]_t)_{t\geq0}$
and 
the corridor-realized vaiance 
$Q^{L,U}(X)=(Q^{L,U}_t(X))_{t\geq0}$
processes 
%which equals the quadratic variation of the Markov chain
%$X$
%and is 
which are for any  maturity
$T$
defined 
%in an analogous way to~\eqref{eq:QuadVarS},
by 
\begin{eqnarray}
\label{eq:QV1MC}
& & [\log(X)]_T   :=   \lim_{n\to\infty}\sum_{t_i^n\in\Pi_n,i\geq1} \left(\log\frac{X_{t_i^n}}{X_{t^n_{i-1}}}\right)^2,\\
& & Q^{L,U}_T(X)  := 
[\log(\olX)]_T-\left(\log\frac{U}{L}\right)^2\sum_{0\leq t\leq T}
\left[\I_{(0,L)}(X_{t-}) \I_{(U,\infty)}(X_{t})+
\I_{(0,L)}(X_{t}) \I_{(U,\infty)}(X_{t-})\right],
\label{eq:QVMarkovChain}
\end{eqnarray}
where partitions 
$\Pi_n$,
$n\in\NN$,
of 
$[0,T]$
are as in~\eqref{eq:QuadVarS},
the process 
$\olX=(\olX_t)_{t\geq0}$
is defined analogously with~\eqref{eq:MinMaxFun}
by
$\olX_t:=\max\{L,\min\{X_t,U\}\}$
%if we substitute 
%upon substituting
%$S_t$
%with
%$X_t$
and
$X_{t-}:=\lim_{s\nearrow t}X_s$
for any
$t>0$.
Note that if we choose 
$L<\min\{x\>:\>x\in E\}$
and
$U>\max\{x\>:\>x\in E\}$,
then the random variables in~\eqref{eq:QV1MC}
and~\eqref{eq:QVMarkovChain}
coincide.
%since in this case the chains
%$\olX$
%and
%$X$
%are the same.
We can therefore without loss of generality
only consider the corridor-realized variance 
$Q^{L,U}_T(X)$.

%where the notation has been  simplified in the following way:
%$t_0=0$,
%$t_N=T$
%for all
%$N\in\NN$
%and the lengths 
%$t_{n}-t_{n-1}$, $n=1,...N$,
%of subintervals in the partition
%go to zero with 
%$N\to\infty$.
%Similarly by replacing the process
%$S$
%%in~\eqref{eq:MinMaxFun}
%and~\eqref{eq:CorrRealVar}
%with the Markov chain 
%$X$
%we can define the corridor-realized variance 
%$Q^{L,U}_T(X)$.

Since the process 
$X$
is a finite-state Markov chain, 
the jumps of 
$X$
arrive with bounded 
%state-dependent 
intensity
and
it is therefore clear that 
the following must hold
\begin{eqnarray}
\label{eq:BasicCondProb}
%& & \PP\left([\log(X)]_{t+\Delta t}-[\log(X)]_t\neq\left(\log\frac{X_{t+\Delta t}}{X_{t}}\right)^2\Bigg\lvert X_t=X(x) \right)
% =  o(\Delta t), 
%\qquad \forall x\in E,\\
& & \PP\left[Q^{L,U}_{t+\Delta t}(X)-Q^{L,U}_{t}(X)\neq\left(\log\frac{\olX_{t+\Delta t}}{\olX_{t}}\right)^2\Bigg\lvert\> X_t=x \right]
 =  o(\Delta t) 
% \quad \forall x\in E\quad
%\forall x\in E\quad \text{such that}\quad 
\quad\text{for all}
\quad x\in[L,U]\cap E.
%\text{if}\quad X_t\in[L,U],
\end{eqnarray}
%for any
%state
%$x\in E$,
An analogous equality holds if 
$X_t$
is ountside of the corridor
$[L,U]$.
Recall also 
that by definition a function 
$f(\Delta t)$
is of the order
$o(\Delta t)$
(usually denoted by
$f(\Delta t)= o(\Delta t)$)
if and only if 
$\lim_{\Delta t\searrow0}f(\Delta t)/\Delta t=0$.
%Here we abuse the notation slightly by denoting
%with 
%$X$
%the deterministic injective function
%$X:E\to \RR_+$
%that maps the abstract state space
%$E$
%into the set of positive real values taken by the asset price
%process %$X$.
Equality~\eqref{eq:BasicCondProb}
implies that the 
$j$-th instantaneous 
conditional moment of the corridor-realized variance
$Q^{L,U}(X)$
is given by
\begin{eqnarray}
M_j(x) & := & \lim_{\Delta t \to0}\frac{1}{\Delta t}
\EE\left[\left(Q^{L,U}_{t+\Delta t}(X)-Q^{L,U}_t(X)\right)^{j}\Big\lvert X_t=x \right] 
\nonumber\\
& = & 
%\I_{[L,U]}(X(x)) \sum_{y\in E} \L(x,y)\left(\log\frac{\olX(y)}{X(x)}\right)^{2j}+
%\I_{(0,L)\cup(U,\infty)}(X(x)) \sum_{y\in E} \I_{[L,U]}(X(y)) \L(x,y)\left(\log\frac{X(y)}{\olX(x)}\right)^{2j}.
%\sum_{y\in E} \L(x,y)\left(\log\frac{X(y)}{X(x)}\right)^{2j}
\sum_{y\in E}\L(x,y)\left[\left(\log\frac{\oly}{\olx}\right)^{2j}
-\left(\log\frac{U}{L}\right)^{2j}
\I_{A_{U,L}}(x,y)
%(\I_{(0,L)}(X(x)) \I_{(U,\infty)}(X(y)) 
%+\I_{(0,L)}(X(y)) \I_{(U,\infty)}(X(x)))
\right]
\label{eq:CondMomentSigma}
\end{eqnarray}
where the set
$A_{U,L}\subset\RR^2$
is defined as
$A_{U,L}:=\left([0,L)\times (U,\infty)\right)
\cup \left((U,\infty)\times[0,L)\right)$
and for any
$x\in E$
we have
$\olx:=\max\{L,\min\{x,U\}\}$.

\subsection{The extension \boldmath $(X,I)$}
The basic idea of this paper is to extend
the markov chain 
$X$
to a 
continuous-time Markov chain
$(X,I)=(X_t,I_t)_{t\geq0}$
where the dynamics
of the process
$I$
approximates well the dynamics of the
corridor-realized variance
$Q^{L,U}(X)$.
Conditional on the path of the chain
$X$,
the process
$I$
will be a compound Poisson process with
jump-intensity that 
is a function 
of the current state
of 
$X$.
%The dynamics of the process
%$I$
%ill be specified by its 
The generator
of
$(X,I)$
will be chosen in such a way that the first
$k\in\NN$
infinitesimal 
moments of 
$I$
and 
$Q^{L,U}(X)$
coincide.
The approximating chain 
$I$ 
will start at 0
(as does the process
$Q^{L,U}(X)$)
%\footnote{This is because $\Sigma_0 = 0$ by definition.} 
and gradually jump 
up its uniform state-space 
$\{0,\alpha,..,\alpha 2C\}$, 
where
$\alpha$ 
is a small positive constant
and
$C$
is
some fixed integer. 

The main computational tool in this paper
is the well-known
spectral decomposition for partial-circulant matrices
(see Appendices A.2-A.4 in~\cite{ALM}
for the definition and the properties 
of the spectrum),
which will be applied to the generator of
the Markov chain
$(X,I)$.
The geometry of the state-space
$\{0,\alpha,..,\alpha 2C\}$ 
%$I(\Psi_C)$
%=\{\alpha a\>:\>a\in\Psi_C\}$
is therefore of fundamental importance
because it allows the generator of
$(X,I)$
to be expressed as a partial-circulant 
matrix.
As mentioned in the introduction,
the main difference between the approach 
in the present paper and the algorithm in~\cite{ALM}
is that here we take advantage of the full strength
of the partial-circulant form of the generator of
$(X,I)$.
This
allows us to define the process 
$I$
as a compound Poisson process with state-dependent intensity
rather than just a Poisson process (which was the case
in~\cite{ALM}),
without adding computational complexity. As we shall see in Section~\ref{sec:numerical}
this enables us to approximate the entire distribution
of the corridor-realized variance and hence
obtian much more accurate numerical
results. 

Assuming that the process
$I$
can jump at most
$n\in\NN$
states up
from its current position
in an infinitesimal amount of time,
the dynamics of 
$I$
are uniquely determined by the state-dependent
intensities 
\begin{eqnarray}
\label{eq:Intensities}
\lambda_i:E\to\RR_+,\quad\text{ where }\quad i\in\{1,\ldots,n\}
\end{eqnarray}
and 
$E$
is the state-space of the chain 
$X$.
The generator of 
$I$,
conditional on the event
$X_t=x$,
%where $x\in E$,
can therefore 
for any
$c,d\in \{0,1,.., 2C\}$ 
be expressed as
\begin{eqnarray}
\label{eq:CondGenOfI}
& & \L^I(x:c,d)\>    := \>    
\left\{
\begin{array}{ll}
\lambda_j(x) & \mathrm{if}\>\>d=c+j \!\!\!\mod (2C+1)\quad \text{for some}\quad j\in\{1,...,n\};\\%\nonumber
-\sum_{i=1}^n\lambda_i(x)  & \mathrm{if}\>\>d=c;  \\
0 & \text{ otherwise.}
\end{array} 
\right.  
\end{eqnarray} 
The dimension of the matrix
$\L^I(x:\cdot,\cdot)$
is
$2C+1$
for all 
$x\in E$
and the identity
$d=c+j \!\!\!\mod (2C+1)$
means that 
the numbers
$d$
and
$c+j$
represent the same element in the additive
group
$\ZZ_{2C+1}$.
%(i.e. the difference
%$d-c-j$
%is divisible by
%$2C+1$).
A key observation here is that the entries 
$\L^I(x:c,d)$
in the conditional generator
depend on 
$c$
and
$d$
solely through the difference 
$d-c$
and hence the 
afore mentioned group structure
makes the conditional generator into a circulant
matrix
(see Appendix~\ref{sec:Bloc}
for the definition of circulant matrices).

This algebraic structure of the conditional 
generator
$\L^I(x:\cdot,\cdot)$
translates into a periodic boundary condition for the process
$I$.
This is very undesirable because the process 
$Q^{L,U}(X)$
we are trying to approximate 
clearly does not exhibit such features.
We must therefore choose 
$C$
large enough so that even if the chain
$I$
is allowed to jump 
$n$
steps up
at any time, the probability that it oversteps the boundary
is negligible
(i.e. below machine precision).
We will see in Section~\ref{sec:numerical}
that in practice
$C\approx 100$
and
$n\approx 30$
is sufficient to avoid the boundary.
Since our aim is to match the first
$k$
instantaneous moments, it is necessary to take
$n$
larger or equal to 
$k$.
In applications this does not pose additional restrictions because, 
as we shall see in 
Section~\ref{sec:numerical},
$k=3$
produces the desired results for jump-diffusions
and 
$k=2$
is already enough for continuous processes.

The conditional generators given in~\eqref{eq:CondGenOfI}
can be used to specify the generator of the Markov chain
$(X,I)$
on the state-space
$E\times\{0,\alpha,\ldots,\alpha2C\}$
as follows
\begin{equation}
\label{eq:lift}
\G(x,c;y,d) := \L(x,y)\delta_{c,d}
+ \L^I(x:c,d)\delta_{x,y},
\end{equation}
where 
$x,y\in E$, 
$c,d\in\{0,1,\ldots,2C\}$ 
and 
$\delta_{\cdot,\cdot}$ 
denotes the
Kronecker delta function.
The matrix 
$\G$
is of the size
$N(2C+1)$
and has partial-circulant form.
In other words  we can express
$\G$
in terms of 
$N^2$
blocks where each block is a square matrix of the size
$2C+1$
and
the blocks that intersects the diagonal 
of 
%the matrix
$\G$
are equal to a sum of a circulant matrix and 
a scalar multiple of the identity matrix.
All other blocks are scalar matrices. For the precise
definition of partial-circulant matrices see Appendix~\ref{sec:Bloc}.

We can now compute,
using~\eqref{eq:CondGenOfI} and~\eqref{eq:lift},
the 
$j$-th 
instantaneous 
conditional 
moment of the process 
$I$
%of the Markov chain
%$(X,I)$
as follows
\begin{eqnarray}
\lim_{\Delta t\rightarrow 0}\frac{1}{\Delta t}
\EE\left[(I_{t+\Delta t}-I_t)^j \Bigl\lvert X_t=x,I_t=\alpha c
 \right]& = & \sum_{d=0}^{2C}(\alpha d-\alpha c)^j\L^I(x:c,d) \nonumber\\
        & = & \alpha^j\sum_{d=1}^n d^j\lambda_d(x) 
\label{eq:Inst_ith_Moment}  
\end{eqnarray}
for any 
$x\in E$
%of the underlying process
%$X_t$
and all integers
$c\in\{0,1,\ldots,2C\}$
that satisfy the inequality
$c<2C-n$,
where 
$n$
was introduced in~\eqref{eq:Intensities}.
This inequality implies that
the process 
$I$
cannot jump to or above
$\alpha 2C$
(i.e. it cannot complete a full circle)
in a very short time interval 
$\Delta t$.
Note also that it is through this inequality only 
that 
identity~\eqref{eq:Inst_ith_Moment}
depends on the current level
$\alpha c$
of the process
$I$.

Our main goal is to approximate the process
$(X,Q^{L,U}(X))$,
where corridor-realized variance 
$Q^{L,U}(X)$
is defined in~\eqref{eq:QVMarkovChain},
by the continuous-time Markov chain
$(X,I)$
with generator given by~\eqref{eq:lift}.
We now match the first 
$k$
instantaneous conditional moments
of processes
$Q^{L,U}(X)$
and
$I$
using 
identities~\eqref{eq:CondMomentSigma}
and~\eqref{eq:Inst_ith_Moment}:
%imply that the following system of equations 
%must be satisfied
\begin{equation}
\label{eq:fundSystem}
\alpha^j\sum_{d=1}^n d^j\lambda_d(x)  =  M_j(x)\quad\text{for any}\quad x\in E\quad\text{and}
\quad j=1,\ldots,k.
\end{equation}
In other words we must choose the intensity functions
$\lambda_i$
(see~\eqref{eq:Intensities})
and the parameter
$\alpha$
so that the system~\eqref{eq:fundSystem}
is satisfied.
The necessary requirement for the solution 
is that 
$\lambda_i(x)\geq0$
for all 
$x\in E$
and all
$i=1,\ldots,n$.
These inequalities can place non-trivial restrictions
on the solution space and will be analysed in more
detail
in Sections~\ref{sec:second_moments}
and~\ref{sec:third_moments}.

Another simple yet important observation that follows 
from~\eqref{eq:fundSystem}
is that, in order to match the first 
$k$
instantaneous conditional moments 
of the corridor-realized variance
$Q^{L,U}(X)$, 
the size of the support of the jump distribution
of the 
of Poisson processes with state-dependant 
intensity
(i.e.
$n$)
must be at least
$k$.
From now on we assume that
$n\geq k$.

The pricing of volatility derivatives is done using
%can be done in this setup
%in the same way as in~ \cite{ALM}.
the following theorem which yields a 
closed-form formula for the semingroup of the Markov chain
$(X,I)$.
\begin{thm}
\label{thm:LiftKer}
Let
$\G$
be the generator matrix of the Markov process
$(X,I)$
given by~\eqref{eq:lift}.
%and let
%$\phi:\CC\rightarrow\CC$
%be a holomorphic function.
Then 
for any
$t\geq0$,
$x,y\in E$
and
$d\in\{0,\ldots,2C\}$
the equality holds
\begin{eqnarray}
\PP\left(X_t=y,I_t=\alpha d\big\lvert X_0=x\right) & = &
\exp(t\G)(x,0;y,d)\nonumber
\\ 
& = & \frac{1}{2C + 1} 
\sum_{j=0}^{2C} e^{\cu p_j d} \exp(t\L_j)(x,y),
\label{eq:joint_pdf}
\end{eqnarray}
where
$\cu =\sqrt{-1}$, 
the scalars
$p_j$
and the 
complex matrices
$\L_j$,
for 
$j=0,\ldots,2C$,
are given by
\begin{eqnarray}
\L_j(x,y) 
&:=& \L(x,y) + \delta_{x,y}
\sum_{i=1}^n 
\left(
e^{-\cu p_j i}-1
\right)
\lambda_i(x),
\label{eq:generator_Ik_new_kth_moment}
\\
p_j & := & \frac{2\pi}{2C+1}j.\nonumber
\end{eqnarray}
%The vectors
%$e_x,e_y$
%denote the corresponding standard basis vectors
%in the euclidean space
%%$\RR^N$
%and $'$
%stands for transposition.
\end{thm}

Theorem~\ref{thm:LiftKer}
is the main computational tool used in this paper
which allows us to find in a
semi-analytic form the
semigroup 
of the chain
$(X,I)$
(if
$C\approx 100$
and
$N=70$,
the matrix
$\G$
contains more than 
$10^8$
elements).
For a straightforward implementation
of the algorithm in Matlab see~\cite{LM_Code}.
It is clear that Theorem~\ref{thm:LiftKer} generalizes equation~(6)
in~\cite{ALM}
and that this generalization involves 
exactly the same number 
of matrix operations 
as the algorithm in \cite{ALM}. 
The only additional computations
are the sums in~\eqref{eq:generator_Ik_new_kth_moment}.
%In fact the special case
%$n=1$
%in
%Theorem~\ref{thm:LiftKer}
%covers the results in~\cite{ALM},
%where only the first instantaneous conditional moment of
%the realized variance was considered. 

%Notice that the generators for $I_t$ are circular matrices which have 
%non-zero entries on the diagonal and $n$ rows above the diagonal. 
%Using expression (8) in \cite{ALM} for the eigenvalues of circulant marices, 
%expression (15) in \cite{ALM} can be reinterpreted as expression (\ref{eq:generator_Ik_new_kth_moment})
%in this paper. 

The proof of Theorem~\ref{thm:LiftKer} 
relies on the partial-circulant
structure of the matrix
$\G$
given in~\eqref{eq:lift}.
The argument follows precisely
the same lines
as the one that proved
Theorem~3.1 in~\cite{ALM}
and will therefore not be given here
(see Appendix~A.5 in~\cite{ALM}
for more details).

Since the dynamics of the process
$(X,I)$
are assumed to be under a risk-neutral measure,
the current value of any payoff that depends on the corridor-realized variance
at fixed maturity can easily be obtained from the formulae 
in Theorem~\ref{thm:LiftKer}.
Furthermore
the same algorithm 
%based on
%Theorem~\ref{thm:LiftKer}
yields the risk sensitivities 
Delta and Gamma of any derivative on the
corridor-realized variance,
without adding computational complexity.
This is 
because the output of our scheme is
%algorithm yields 
a vector of values
of the derivative in question conditional on 
the process
$X$
starting at 
each of the elements in its state-space. 
We should also note that forward-starting derivatives on
the corridor-realized variance can be dealt with using the same algorithm 
because conditioning on the state of a Markov chain at a future time
requires only a single additional matrix-vector multiplication.
Explicit calculations are obvious and are omitted
(see~\cite{ALM} for more details).

\section{Convergence}
\label{sec:Conv}
In Section~\ref{sec:k_moments}
we defined the Markov chain
$(X,I^k)$
via its generator~\eqref{eq:lift}
that in some sense approximates the process
$(X,Q^{L,U}(X))$,
where 
$Q^{L,U}(X)$
is the corridor-realized variance of 
$X$
defined in~\eqref{eq:QVMarkovChain}.
Here
$I^k$
denotes the process
$I$
from 
Section~\ref{sec:k_moments}
which satisfies the instantaneous conditional
moment restrictions, given by~\eqref{eq:fundSystem},
up to order
$k$.

Notice that it follows directly from definition~\eqref{eq:QVMarkovChain} 
that the process
$(X,Q^{L,U}(X))$
is adapted to the natural filtration generated by the 
chain
$X$
and that its components 
$X$
and
$Q^{L,U}(X)$
can only jump simultaneously.
On the other hand note that the form of the generator of the chain
$(X,I^k)$,
given by~\eqref{eq:lift},
implies that the components
$X$
and
$I^k$
cannot both jump at the same time.
It is also clear that the process
$I^k$
is not adapted to the natural filtration of 
$X$.
In this section our goal is to prove that, in spite of these differences,
for any fixed time 
$T$
the sequence of random variables
$(I^k_T)_{k\in\NN}$
converges in distribution to the random variable 
$Q^{L,U}_T(X)$.
In fact we have the following theorem
which 
states that, for any bounded European payoff, the price of 
the corresponding derivative on the corridor-realized variance
in the approximate model
$(X,I^k)$
converges to the price of the same derivative in
$(X,Q^{L,U}(X))$
as the number 
$k$
of matched instantaneous conditional moments
tends to infinity.

\begin{thm}
\label{thm:Conv}
Let
$X$
be a continuous-time Markov chain with generator
$\L$
as given in Section~\ref{sec:k_moments}.
For each 
$k\in\NN$
define a real number
\begin{eqnarray}
\label{eq:CondFor_k}
\alpha_k & :=  & \frac{1}{k}\max\left\{\left(\log\frac{y}{x}\right)^2\>:\>x,y\in E\backslash \{0\}\right\},
%\quad\text{where}\quad X:E\to(0,\infty),
%\quad\text{where}\quad M:=\max\left\{\log\left(\frac{X(y)}{X(x)}\right)^2\>:\>x,y\in E\right\},
\end{eqnarray}
assume that 
$n$
in~\eqref{eq:Intensities}
equals 
$k$
and that there exist 
functions
$\lambda_i^k:E\to\RR_+,$
$i\in\{1,\ldots,k\}$,
that solve the system of equations~\eqref{eq:fundSystem}.
Let the continuous-time Markov chain
$(X,I^k)$ be given by generator~\eqref{eq:lift}
where 
%integer 
%$n$
%in~\eqref{eq:Intensities}
%equals 
%$k$
%and the 
%functions
%$\lambda_i^k:E\to(0,\infty),$
%$i\in\{1,\ldots,k\}$,
%are assumed to solve the system of equations
%with
%$\alpha:=\alpha_k$
%specified by the condition
%\begin{eqnarray}
%\label{eq:CondFor_k}
%%k\alpha_k & =  & \max\left\{\log\left(\frac{X(y)}{X(x)}\right)^2\>:\>x,y\in E\right\}.
%\end{eqnarray}
%We also assume that 
the integers
$C_k$
in~\eqref{eq:CondGenOfI},
which determine the size of the state-space of the process
$I^k$,
are chosen in such a way that
$\lim_{k\to\infty} \alpha_k C_k=\infty$.
Then for any fixed time
$T>0$
the sequence of random  variables $(I_T^k)_{k\in\NN}$
converges weakly to 
$Q^{L,U}_T(X)$.
In other words
for any bounded continuous function
$f:\RR\to\RR$
we have
%$s\in\RR_+$
%at which the cumulative distribution function 
%$x\mapsto \PP(\Sigma_T\leq x)$
%is continuous
%we have
%$$\lim_{k\to\infty}\PP(I_T^k\leq s)=\PP(\Sigma_T\leq s).$$
$$\lim_{k\to\infty}\EE[f(I_T^k)\lvert X_0]=\EE[f(Q^{L,U}_T(X))\lvert X_0].$$
\end{thm}

Before proving Theorem~\ref{thm:Conv}
we note that the assumption on the existence of non-negative 
solutions of the system in~\eqref{eq:fundSystem}
is not stringent and can be satisfies
for any chain 
$X$
by allowing 
$n$
in~\eqref{eq:Intensities}
to take values larger than 
$k$.
The restriction 
$n=k$
in Theorem~\ref{thm:Conv}
is used because it simplifies the notation.

\begin{pr}
Throughout this proof we will use the notation 
$\Sigma_t:=Q^{L,U}_t(X)$
for 
any
$t\in\RR_+$.
By the L\'{e}vy continuity theorem it is enough to prove that the equality 
holds
$$\lim_{k\to\infty}\EE[\exp(\cu u I_T^k)]=\EE[\exp(\cu u \Sigma_T)]\quad\text{for each}\quad u\in\RR.$$
Let
$\Delta t>0$
be a small positive number and note that, by conditioning on the 
$\sigma$-algebra generated by the process 
$X$
up to and including time
$T-\Delta t$
and using the Markov property,
we obtain the following representation
\begin{eqnarray}
\label{eq:SigmaBound}
\EE[\exp(\cu u \Sigma_T)]  & =  & 
\EE\left[ \exp(\cu u \Sigma_{T-\Delta t})
\EE\left[\exp(\cu u(\Sigma_T- \Sigma_{T-\Delta t}))\big\lvert X_{T-\Delta t} \right]\right]\\
& = & 
\EE\left[ e^{\cu u \Sigma_{T-\Delta t}} \left(\sum_{j=0}^k\frac{(\cu u)^j}{j!}
\EE\left[(\Sigma_T- \Sigma_{T-\Delta t})^j\big\lvert X_{T-\Delta t} \right]\right.\right. \nonumber\\
& &\left.\left. +\sum_{j=k+1}^\infty\frac{(\cu u)^j}{j!}
\EE\left[(\Sigma_T- \Sigma_{T-\Delta t})^j\big\lvert X_{T-\Delta t} \right]\right)\right]\nonumber\\
& = & 
\EE\left[e^{\cu u \Sigma_{T-\Delta t}} \left(1+\Delta t \sum_{j=1}^k\frac{(\cu u)^j}{j!}
M_j(X_{T-\Delta t})\right.\right. \nonumber\\
& & \left.\left. +\sum_{j=k+1}^\infty\frac{(\cu u)^j}{j!}
\EE\left[(\Sigma_T- \Sigma_{T-\Delta t})^j\big\lvert X_{T-\Delta t}\right]\right) \right]+o(\Delta t),\nonumber
\end{eqnarray}
where 
$M_j$
is defined in~\eqref{eq:CondMomentSigma}.
By applying Markov property of 
$(X,I^k)$,
identity~\eqref{eq:Inst_ith_Moment} 
and condition~\eqref{eq:fundSystem},
which holds by assumption for all
$j\in\{1,\ldots,k\}$,
we obtain
\begin{eqnarray}
\label{eq:IBound} 
\EE[\exp(\cu u I^k_T)]  & =  & 
\EE\left[e^{\cu u I^k_{T-\Delta t}} \left(1+\Delta t \sum_{j=1}^k\frac{(\cu u)^j}{j!}
M_j(X_{T-\Delta t})\right.\right. \\
& & \left.\left. +\sum_{j=k+1}^\infty\frac{(\cu u)^j}{j!}
\EE\left[(I^k_T- I^k_{T-\Delta t})^j\bigg\lvert X_{T-\Delta t},I^k_{T-\Delta t}\right]\right) \right]+o(\Delta t).
\nonumber
\end{eqnarray}

It follows from~\eqref{eq:CondMomentSigma}
that there exists a positive constant
$G$
such that 
$\max\{M_j(x)\>:\> x\in E\}\leq G^j$
for all
$j\in\NN$.
Therefore we find 
that 
for a constant 
$D:=\exp(uG)$
the following
inequality
holds 
on the entire probability space
\begin{eqnarray}
\label{eq:firstBound}
\bigg\lvert \sum_{j=1}^k\frac{(\cu u)^j}{j!}
M_j(X_{T-\Delta t})\bigg\rvert \leq D.
\end{eqnarray}
Note also that 
$D$
is independent of 
$k$
and
$\Delta t$.

Definition~\eqref{eq:CondFor_k}
implies that 
$k\alpha_k$
is a positive constant,
say
$A$,
for each 
$k\in\NN$.
If we introduce a positive constant 
$L:=\max\{-\L(x,x)\>:\>x\in E\}$,
we obtain
%, using
%equality~\eqref{eq:BasicCondProb},
the following bound 
\begin{eqnarray}
\label{eq:Large_n_BoundSigma}
\EE\left[(\Sigma_T- \Sigma_{T-\Delta t})^j\big\lvert X_{T-\Delta t}\right] & \leq & 
A^j L \Delta t +o(\Delta t)\quad\text{for each}\quad j\in\NN
\end{eqnarray}
on the entire probability space. 
In order to find a similar bound for the process
$I^k$
we first note that if follows from the linear equation~\eqref{eq:fundSystem}
(for 
$j=1$)
and definition~\eqref{eq:CondFor_k}
that the inequalities
$$\sum_{d=1}^k d\lambda_d^k(x)\leq k L
%\max\{\lambda_i^k(x)\>:\>x\in E\}\leq L,
\quad\text{for all}\quad k\in\NN,\>x\in E
%\quad i\in\{1,\ldots,k\},
$$
must hold.
Therefore~\eqref{eq:Inst_ith_Moment} 
%and
%an argument similar to the one that 
%yielded~\eqref{eq:Large_n_BoundSigma}
implies
\begin{eqnarray}
\label{eq:Large_n_BoundI}
\EE\left[(I^k_T- I^k_{T-\Delta t})^j\bigg\lvert X_{T-\Delta t},I^k_{T-\Delta t}\right]
%& \leq & (k\alpha_k)^n +o(\Delta t),\\
%\EE\left[(\Sigma_T- \Sigma_{T-\Delta t})^n\big\lvert X_{T-\Delta t}\right]\right\} 
& \leq & 
%\sum_{d=1}^k (\alpha_k d)^j\lambda_d^k(x)\leq 
%A^j \sum_{d=1}^k \lambda_d^k(x)\leq 
A^j k L \Delta t +o(\Delta t)\quad\text{for any}\quad j\in\NN
%A^j +o(\Delta t)
\end{eqnarray}
and any small time-step
$\Delta t$.
We can now combine the estimates 
in~\eqref{eq:SigmaBound}, \eqref{eq:IBound},
\eqref{eq:firstBound}, \eqref{eq:Large_n_BoundSigma}
and~\eqref{eq:Large_n_BoundI}
to obtain the key bound
\begin{eqnarray}
\label{eq:KeyBound}
\left\lvert\EE[\exp(\cu u \Sigma_T)] - \EE[\exp(\cu u I^k_T)]\right\rvert  & \leq  & 
%\left\lvert\EE[e^{\cu u \Sigma_T}] - \EE[e^{\cu u I^k_T}]\right\rvert  & \leq  & 
\left\lvert\EE[\exp(\cu u \Sigma_{T-\Delta t})] - \EE[\exp(\cu u I^k_{T-\Delta t})]\right\rvert(1+\Delta t D)\\
%\left\lvert\EE[e^{\cu u \Sigma_{T-\Delta t}}] - \EE[e^{\cu u I^k_{T-\Delta t}}]\right\rvert(1+D\Delta t)
& & +\>\> L (k+1)\Delta t \sum_{j=k+1}^\infty\frac{(Au)^j}{j!}+o(\Delta t).\nonumber
\end{eqnarray}

The main idea of the proof of Theorem~\ref{thm:Conv}
is to iterate the bound in~\eqref{eq:KeyBound}
$\frac{T}{\Delta t}$
times. This procedure yields the following estimates
\begin{eqnarray*}
\left\lvert\EE[\exp(\cu u \Sigma_T)] - \EE[\exp(\cu u I^k_T)]\right\rvert  & \leq  & 
%\left\lvert\EE[e^{\cu u \Sigma_T}] - \EE[e^{\cu u I^k_T}]\right\rvert  & \leq  & 
D\Delta t(1+\Delta t D)^{(T/\Delta t)-1}+
L (k+1)T \sum_{j=k+1}^\infty\frac{(Au)^j}{j!}+T\frac{o(\Delta t)}{\Delta t}.
\end{eqnarray*}
%where
%$L'$
%is some positive constant.
Since the left-hand side of this inequality is independent of 
$\Delta t$,
the inequality must hold in the limit
as
$\Delta t\searrow0$.
We therefore find
\begin{eqnarray}
\label{eq:inequalityFinal}
\left\lvert\EE[\exp(\cu u \Sigma_T)] - \EE[\exp(\cu u I^k_T)]\right\rvert   \leq   
L (k+1)T \sum_{j=k+1}^\infty\frac{(Au)^j}{j!}.
\end{eqnarray}
The right-hand side of inequality~\eqref{eq:inequalityFinal}
clearly converges to zero 
as 
$k$
tends to infinity.
This concludes 
the proof of the theorem.
\end{pr}

Theorem~\ref{thm:Conv}
implies that the prices of the volatility derivatives in the 
Markov chain model 
$X$
can be 
approximated arbitrarily well using the method defined in Section~\ref{sec:k_moments}.
Our initial problem of approximating prices in the model based on a continuous-time 
Markov process
$S$
is 
by Theorem~\ref{thm:Conv} reduced to the question 
of the approximation of the law of 
$S$
by the law of
$X$.
This can be achieved by a judicious
choice for the generator matrix of the chain
$X$.
Since this is not the central topic of this paper
we will not investigate the question further in this 
generality
(see~\cite{Kurtz} for numerous results 
on weak convergence of Markov processes).
However
in Sections~\ref{sec:second_moments}
and~\ref{sec:third_moments}
we are going to propose specific 
Markov chain approximations for
diffusion and jump-diffusion processes 
respectively and study numerically the behaviour of the
approximations for volatility derivatives 
in Section~\ref{sec:numerical}.

\section{The realized variance of a diffusion process}
\label{sec:second_moments}
Our task now is
to apply the method described in Section~\ref{sec:k_moments}  
to approximate the dynamics of the corridor-realized variance of a diffusion processes.
The first step is to approximate the diffusion process 
$S$
which solves  
the stochastic differential equation (SDE)
\begin{eqnarray}
\label{eq:SDE}
\frac{dS_t}{S_t} =  \gamma dt+\sigma\left(\frac{S_t}{S_0}\right)dW_t,%\qquad\text{with measurable volatility function}\quad
%\sigma:\RR_+\to\RR_+,
%is a measurable function,
\end{eqnarray}
with measurable volatility function
$\sigma:\RR_+\to\RR_+$,
using a continuous-time Markov chain
$X$.
A possible way of achieving this is to use a 
generator for the chain 
$X$
given by the following system of linear equations
\begin{eqnarray}
\label{eq:GenEqu}
\sum_{y\in E} \L(x,y) &=& 0,\nonumber \\
\sum_{y\in E} \L(x,y) (y-x) &=& \gamma x, \\
\sum_{y\in E} \L(x,y) (y-x)^2 &=&  \sigma\left(\frac{x}{X_0}\right)^2x^2\nonumber 
\end{eqnarray}
for each
$x\in E$.
In Appendix~\ref{sec:NU_State_Space}
we give an algorithm to define the state-space
$E$
of the chain
$X$.
In Section~\ref{sec:numerical}
we provide a numerical comparison for vanilla option prices
in the CEV model,
i.e.
in the case
$\sigma(s):=\sigma_0 s^{\beta-1}$,
and in the corresponding Markov chain model 
given by the approximation above.
Note that
a Markov chain approximation 
$X$
of the diffusion 
$S$
is in the spirit of~\cite{AM}
and is by no means the only viable alternative. One could
produce more accurate results by matching higher instantaneous
moments of the two processes
(see~\cite{Mijatovic} for rates of convergence in some special cases).

If the solution of SDE~\eqref{eq:SDE}
is used as a model for the risky security 
under a risk-neutral measure
we have to stipulate that
$\gamma=r$,
where 
$r$
is the prevailing 
risk-free rate in the economy.
Therefore by the first two equations in system~\eqref{eq:GenEqu}
the vector 
in
$\RR^N$
with cooridnates equal to the
elements in the set
$E$
%$\sum_{x\in E}xe_x$
%viewed as a vector in a euclidean space
represents an eigenvector of the matrix
$\L$
for the eigenvalue 
$\gamma$.
Hence we find
\begin{eqnarray}
\label{eq:DriftX}
\EE[X_t\lvert X_0=x]=e_x'\exp(t\L)\sum_{y\in E}ye_y %=e^{t\gamma}e_x'X
& = & e^{t\gamma}x,\qquad\forall x\in E,
\end{eqnarray}
where
$e_x$
denotes the standard basis vector in 
$\RR^N$
that corresponds to the element 
$x\in E$
in the natural ordering
and the operation
$'$
denotes transposition.
Therefore,
under the condition
$\gamma=r$,
the market driven by the chain 
$X$ 
will also have a correct risk-neutral drift.

Once we define the chain 
$X$,
the next task is to specify the process
$I$
that approximates well the corridor-realized variance 
$Q^{L,U}(X)$
defined in~\eqref{eq:QVMarkovChain}.
As we shall see in Section~\ref{sec:numerical},
matching the first two moments 
(i.e. the case 
$k=2$
in Section~\ref{sec:k_moments}) 
is sufficient to approximate the corridor-realized variance dynamics
of a diffusion processes.
It is therefore necessary to take 
$n\geq2$,
where 
$n$
is the number of states the approximate variance
process 
$I$
can jump up by at any given time
(see~\eqref{eq:Intensities}).
To have flexibility we use
$n$
much larger than 
2,
usually around 30.
However 
in order to maintain the tractability of
the solution of system~\eqref{eq:fundSystem}
we make an additional assumption 
that the intensities
$\lambda_i$
in~\eqref{eq:Intensities},
for 
$i=2,\ldots,n$,
%coincide and 
are all equal to a single intensity function
$\lambda_n:E\to\RR_+$.
To simplify the notation we introduce the symbol
\begin{eqnarray}
\label{eq:Symbol}
b_j^{n,m} & := & \sum_{l=n+1}^m l^j,\qquad\text{where}\quad j,n,m\in\NN\quad\text{and}\quad m>n.
\end{eqnarray}
System~\eqref{eq:fundSystem}
%consists of
%
%Assuming that
%the constant grid spacing 
%$\alpha$ 
%is chosen and the maximum jump size of $I_t$ is $n\alpha$
%where $n$ is a positive integer. If $n>2$ then we would have more degrees of freedom than the equations that we are trying to 
%satisfied (i.e. matching the first and second instantaneous moments). We made the choice to set the intensities correspond to the jump size $2\alpha$ 
%to $n\alpha$ to be the same. The intensities
%correspond to $\alpha$ and $2\alpha$ to $n\alpha$ are denoted as $\lambda_1$
%and $\lambda_n$ respectively.
%Conditioning on a level of the underlying (let's say $X(a)$), the intensities are obtained
%by solving the following simultaneous equations:
can in this case be solved explicitly 
as follows
%\begin{eqnarray*}
%\lambda_1(x) 
%+ \lambda_n(x) \sum_{i=2}^n i &=& M_1(x)/\alpha,\\
% \lambda_1(x) 
%+ \lambda_n(x)\sum_{i=2}^n i^2  &=& M_2(x)/\alpha^2.
%\end{eqnarray*}
%After some algebras, the intensities can be computed explicitly as
\begin{eqnarray}
\label{eq:l1Diff}
\lambda_1(x) &=& \frac{ \alpha M_1(x)b_2^{1,n} - M_2(x) b_1^{1,n}}
                      {\alpha^2(b_2^{1,n}-b_1^{1,n})},\quad \text{for any}\quad x\in E, \\ 
\lambda_n(x) &=& \frac{M_2(x)-\alpha M_1(x)}
                      {\alpha^2(b_2^{1,n}-b_1^{1,n})},\quad \text{for any}\quad x\in E, 
\label{eq:lnDiff}
\end{eqnarray}
where 
$M_j(x)$
is given in~\eqref{eq:CondMomentSigma}.
%As mentioned in Sectio~\ref{sec:k_moments}
%the 
Since the functions
$\lambda_1,\lambda_n$
%in~\eqref{eq:Intensities}
are intensities,  
all the values they take must be
non-negative. 
The formulae above imply that this
is satisfied if and only if the following inequalities hold
\begin{eqnarray}
\alpha\frac{b_2^{1,n}}{b_1^{1,n}}\>\geq\> \frac{M_2(x)}{M_1(x)} \>
\geq\> \alpha \qquad
\text{for all}\quad x\in E.
\label{eq:FindAlphaDiff}
\end{eqnarray}
It is clear that the function
$x\mapsto M_2(x)/M_1(x)$,
$x\in E$,
depends on the definition of the chain 
$X$
both through the choice of the state-space
$E$
and the choice of the generator
$\L$.
Figure~\ref{fig:cev_VQ_ratio}
contains the plot of this function
in the 
special case of the CEV model. 
Inequalities~\eqref{eq:FindAlphaDiff}
are used to help us choose a feasible value 
for the parameter
$\alpha$
which determines the geometry of the 
state-space of the process 
$I$.
Note also 
that~\eqref{eq:FindAlphaDiff}
implies that 
the larger the value of 
$n$
is,
the 
less restricted we are
in choosing 
$\alpha$.
In Section~\ref{sec:numerical}
we will make these choices explicit for the 
CEV model.
%make the second condition less stringent.
%imply  that
%the parameter
%$\alpha$
%(i.e. the spacing of the lattice for the process
%$I$)
%must be chosen in such a way that it
%satisfies
%$$
%\alpha\leq \min_{x\in E}\frac{M_2(x)}{M_1(x)}\quad\text{and}\quad
%\frac{b_1^{1,n}}{b_2^{1,n}} \max_{x\in E}\frac{M_2(x)}{M_1(x)}
%\leq \alpha.
%$$
%which implies the followings:
%\begin{itemize}
%\item The lattice spacing $\alpha$ must be less than $\min_{a\in\Omega}\frac{M_2(a)}{M_1(a)}$.
%%\item The biggest jump size multiple $n$ must be chosen such that $\alpha\left(\frac{\sum_{i=2}^n i^2}{\sum_{i=2}^n i}\right)$ is larger than $\max_{a\in\Omega}\frac{M_2(a)}{M_1(a)}$.
%\end{itemize}
%In other words, the instantaneous first and second moment of the realized
%variance $\Sigma_t$ determine the structure of the approximating
%Markov chain $I_t$.

The generator of the approximate corridor-realized variance
$I$, 
conditional on the chain
$X$
being at the level 
$x$, 
is in general 
given by Formula~\eqref{eq:CondGenOfI}.
In this particular case 
the non-zero matrix elements 
$\L^I(x:c,d)$,
$c,d\in \{0,1,\ldots,2C\}$,
are given by
\begin{eqnarray*}
\L^I(x:c,d) & := &
%\L^I(x:c,d) := 
\left\{
\begin{array}{ll}
 \lambda_1(x) & \mathrm{if}\>\>d=(c+1)\!\!\!\mod (2C+1);\\%\nonumber
 \lambda_n(x) & \mathrm{if}\>\>d=(c+i)\!\!\! \mod (2C+1),\> i\in\{2,...,n\};\\               
-\lambda_1(x)-(n-1)\lambda_n(x)  & \mathrm{if}\>\>d=c. %\!\!\!  \mod (2C+1),
\end{array}
\right. 
\end{eqnarray*}
%$\L^I(x:c,d) := \lambda_n(x)$
%if integer
%$((d-c)\mod(2C+1))$
%is in the set
%$\{2,...,n\}$.
This defines explicitly
(via equations~\eqref{eq:l1Diff} and~\eqref{eq:lnDiff})
the dynamics of the chain
$(X,I)$
if the original asset price process 
$S$
is a diffusion.
In Section~\ref{sec:numerical}
we will describe an implementation of this method 
when 
$S$
follows
a CEV process
and study the behaviour of certain 
volatility derivatives
in this model.

\begin{comment}
These choices will be studied numerically in case 
of the CEV model in Section~\ref{sec:numerical}.
and 
is therefore of the form

and the expression (\ref{eq:generator_Ik_new_kth_moment})
can be written explicitly as
\begin{eqnarray}\label{eq:generator_Ik_new_second_moment}
\L_k(a,b) 
&:=& \L(a,b) + \delta_{a,b}
\left(
\lambda_0(a)
+
\lambda_1(a)e^{-\i p_k}
+
\lambda_n(a)\sum_{i=2}^n e^{-\i p_k i}
\right),
\\\nonumber
\end{eqnarray}
where
$\lambda_0(a)=-\lambda_1(a)-(n-1)\lambda_n(a).$
\end{comment}

\section{The realized variance of a jump-diffusion}
\label{sec:third_moments}
In this section the task is to describe the algorithm for the pricing 
of volatility derivatives in jump-diffusion models.
This will be achieved by an application of the algorithm from
Section~\ref{sec:k_moments}
with 
$k=3$.
In Section~\ref{sec:numerical}
we will investigate numerically 
the quality of this approximation. 
We start by describing a construction of the
Markov chain which is used to approximate a jump-diffusion.

\subsection{Markov chain approximations for jump-diffusions}
\label{subsec:Jump_Diff}

We will consider a class of processes with jumps 
that is obtained by subordination 
of diffusions.
The prototype for such processes is 
the well-known variance gamma model
defined in~\cite{MCC},
which can be expressed as 
a time-changed Brownian motion with drift. 

A general way of building (possibly infinite-activity)
jump-diffusion processes is by subordinating
diffusions using a class of independent stochastic time
changes.
Such a time change is given by a non-decreasing
stationary process
$(T_t)_{t\geq0}$
with independent increments,
which starts at zero, 
%The time change
%$T_t$
and is known as a
\textit{subordinator}.
The law 
of
$(T_t)_{t\geq0}$
is characterized by the  
\textit{Bernstein function}
$\phi(\lambda)$,
defined by the following identity
\begin{eqnarray}
\label{eq:BernstienIdentity}
 \EE\left[\exp(-\lambda T_t) \right] = \exp(-\phi(\lambda)t)\quad\text{for any}\quad t\geq0\quad\text{and}
\quad \lambda\in D,
\end{eqnarray}
where 
$D$
is an interval in 
$\RR$
that contains the half-axis
$[0,\infty)$.
%In other words the process
%$T_t$
%is a non-decreasing stationary process
%with independent increments whose Laplace transform
%takes the special form
%$e^{-\phi(\lambda)t}$,
%where
%$\phi(\lambda)$
%is the Bernstein function of the process.
For example in the case of the variance gamma process,
the Bernstein function is of the form
\begin{eqnarray}
\label{bochner}
\phi(\lambda) =
\frac{\mu^2}{\nu}\log\left(1+\lambda\frac{\nu}{\mu}\right).
\end{eqnarray}
In this case 
$(T_t)_{t\geq0}$
is a gamma process\footnote{
The parameter
$\mu$
is the mean rate,
usually taken to be equal 
to one
in order to ensure that
$\EE[T_t]=t$
for all
$t\geq0$, 
and
$\nu$
is the variance rate
of
$(T_t)_{t\geq0}$.} 
with characteristic function equal to
$\EE[\exp(\cu uT_t)]=\exp(-\phi(-\cu u)t)$.
Note that the set 
$D$
in~\eqref{eq:BernstienIdentity}
is in this case equal to
$(-\mu/\nu,\infty)$
(see~\cite{MCC}, equation~(2)).
This subordinator is 
used to construct 
the 
jump-diffusions in
Section~\ref{sec:numerical}. 

Let 
$S$
be a diffusion defined by the SDE in~\eqref{eq:SDE}.
If we evaluate the process 
$S$
at an independent 
subordinator
$(T_t)_{t\geq0}$,
we obtain a Markov process with jumps
$(S_{T_t})_{t\geq0}$.
It was shown in~\cite{Phillips}
that the semigroup of 
$(S_{T_t})_{t\geq0}$
is generated by the unbounded differential 
operator
$\G':=-\phi(-\G)$,
where
$\G$
denotes the generator of the diffusion
$S$.
Similarly, if 
$X$
is a continuous-time Markov chain with 
generator 
$\L$
defined in the first paragraph of Section~\ref{sec:second_moments},
the subordinated process
$(X_{T_t})_{t\geq0}$
is again a continuous-time Markov chain with the generator matrix
$\L':=-\phi(-\L)$.
We should stress here that 
it is possible to define rigorously 
the operator 
$\G'$
using
the spectral decomposition of %the generator
$\G$
and 
%an application of 
the theorey of functional calculus 
(see~\cite{DunfordS}, Chapter~XIII, Section~5, Theorem~1).
The matrix 
$\L'$
can be defined and calculated easily using the Jordan decomposition
of the generator
$\L$.
If the matrix 
$\L$
%is equivalent to a diagonal matrix
%$\Lambda$
can be expressed in the diagonal form
$\L=U\Lambda U^{-1}$,
which is the case in any practical application
(the set of matrices that cannot be diagonalised 
is of codimention one in the space of all matrices
and therefore has Lebesgue measure zero), 
we can compute
$\L'$
using the following formula
\begin{eqnarray}
\label{eq:SubDiag}
\L'  = - U\phi(-\Lambda) U^{-1}.
\end{eqnarray}
Here
$\phi(-\Lambda)$
denotes a
diagonal matrix with diagonal elements
of the form
$\phi(-\lambda)$,
where
$\lambda$
runs over the spectrum
of the generator
$\L$.

Before using the described procedure to define the jump-diffusion process,
we have to make sure that it has the correct drift under a risk-neutral
measure. 
Recall that if the process
$S$
solves the
SDE 
in~\eqref{eq:SDE},
then the identity
$\EE[S_t\lvert S_0]=S_0 \exp(t\gamma)$
holds,
where
$\gamma$
is the drift parameter
in~\eqref{eq:SDE}.
Since 
the subordinator 
$(T_t)_{t\geq0}$
is independent of 
$S$,
by conditioning
on the random variable 
$T_t$,
we find that under the pricing measure
the following identity must hold
$$S_0\exp(rt) = \EE[S_{T_t}\lvert S_0]=S_0 \EE[\exp(\gamma T_t)]=S_0\exp(-\phi(-\gamma)t),$$
where
$\phi$
is the Bernstein function of the subordinator
$(T_t)_{t\geq0}$
and
$r$
is the prevailing risk-free rate which is assumed to be constant.
This will be satisfied if and only if
$r=-\phi(-\gamma)$,
which in case of the gamma subordinator (i.e. when the function 
$\phi$
is given by~\eqref{bochner})
yields an explicit formula for the drift in equation~\eqref{eq:SDE}
\begin{eqnarray}
\label{eq:DriftSub}
\gamma & = & \frac{\mu}{\nu}\left(1-\exp\left(-\frac{r\nu}{\mu^2}\right)\right).
\end{eqnarray}
Since 
formula~\eqref{eq:DriftX}
holds for the chain 
$X$,
tower property 
%for conditional expectations 
and the identity
$r=-\phi(-\gamma)$
imply
$$\EE[X_{T_t}\lvert X_0]=X_0\EE[\exp(\gamma T_t)]=X_0\exp(rt).$$
Therefore the subordinated Markov chain 
$(X_{T_t})_{t\geq0}$
can also be used as a model for a risky asset under 
the pricing measure.

The construction of jump-diffusions described above is
convenient because we can use the generator
$\L$,
that was defined in Section~\ref{sec:second_moments},
and apply the 
Bernstein function 
$\phi$
from~\eqref{bochner}
to obtain the generator of the Markov chain that approximates
the process 
$(S_{T_t})_{t\geq0}$.
This accomplishes the first 
step in the approximation scheme outlined 
in the introduction.
%at the beginning of
%Section~\ref{sec:k_moments}.
In Subsection~\ref{subsec:JumpAlG} we
develop an algorithm for computing the law of the 
relized variance of the approximating chain generated 
by
$\L'$.
It should be stressed that the algorithm in the next subsection
does not 
depend on the procedure used to 
obtain the generator of the approximating chain.

\subsection{The algorithm}
\label{subsec:JumpAlG}
To simplify the notation let us assume that 
$S$
is a jump-diffusion and that 
$X$
is a coninuous-time Markov chain with generator
$\L$
that is used to approximate the dynamics of 
the Markov process
$S$.
Since 
$S$
has jumps 
it is no longer enough to 
use the algorithm from
Section~\ref{sec:k_moments}
with 
$k=2$
(this 
will become clear from the numerical 
results in Section~\ref{sec:numerical}).
In this 
subsection we give an account 
of how to apply our algorithm
%form
%Section~\ref{sec:k_moments}
in the case
$k=3$.

Assume we have chosen
the spacing 
$\alpha$ 
and the constant
$C$
that uniquely determine the geometry
of the state-space 
of the process
$I$
(see the paragraph preceding equation~\eqref{eq:Intensities}
for the definition of the state-space).
Set the maximum jump size of
$I$
to be
$m\alpha$
for some
$m\in\NN$.
We now pick an integer
$n$,
such that 
$1<n<m$,
%Our goal is to match $M_i(a),\>i=1,2,3$ for every $a\in\Omega$ with
%a state-dependent compound Poisson process $I_t$. 
and set the intensities that correspond 
to the jumps of the process
$I$
of sizes between
$2\alpha$ 
and 
$n\alpha$ 
to equal
$\lambda_n$.
%equal to a function
%$\lambda_n:E\to(0,\infty)$.
Similarley we set the intensities for the jumps of
sizes 
between
$(n+1)\alpha$ 
and
$m\alpha$ 
to be equal to 
$\lambda_m$.
%be given by the function
%$\lambda_m:E\to(0,\infty)$.
%with $n\leq m$. 
This simplifying assumption makes it possible to describe 
the dynamics of 
$I$
using only three functions
$\lambda_1, \lambda_n, \lambda_m:E\to\RR_+$
that give state-dependent intensities for jumping up
by 
$i\alpha$ 
where
$i=1$,
$i\in\{2,\ldots,n\}$,
$i\in\{n+1,\ldots,m\}$
respectively.
In order to match 
$k=3$
instantaneous conditional moments of the 
corridor-realized variance
$Q^{L,U}(X)$,
these functions must
by~\eqref{eq:fundSystem}
satisfy the following system of equations
\begin{eqnarray*}
\begin{pmatrix}
1 & b_1^{1,n} & b_1^{n,m}\\
1 & b_2^{1,n} & b_2^{n,m}\\
1 & b_3^{1,n} & b_3^{n,m}
\end{pmatrix}
\begin{pmatrix}
\lambda_1(x)\\
\lambda_n(x)\\
\lambda_m(x)
\end{pmatrix} & = &
\begin{pmatrix}
\overline{M}_1(x)\\
\overline{M}_2(x)\\
\overline{M}_3(x)
\end{pmatrix}\quad 
\forall x\in E,
%\lambda_1(x) + b_j^{1,n}\lambda_n(x) + b_j^{n,m}\lambda_m(x) & = &\overline{M}_j(x) 
%\quad\text{for all}\quad 
%\quad\text{and}\quad j=1,2,3,
\quad\text{where}\quad
\overline{M}_j(x)  := \frac{M_j(x)}{\alpha^j},\quad
\end{eqnarray*}
the symbol 
$b_j^{n,m}$
is defined in~\eqref{eq:Symbol}
and functions
$M_j$,
$j=1,2,3$,
are given in~\eqref{eq:CondMomentSigma}.
Gaussian elimination yields the explicit solution of the system
\label{SystemSol}
\begin{eqnarray}
\lambda_1 &=& \frac{(\overline{M}_3b_1^{n,m}-\overline{M}_1b_3^{n,m})(b_2^{1,n}b_1^{n,m}-b_1^{1,n}b_2^{n,m})-(\overline{M}_2b_1^{n,m}-\overline{M}_1b_2^{n,m})(b_3^{1,n}b_1^{n,m}-b_1^{1,n}b_3^{n,m})}
                   {(b_1^{n,m}-b_3^{n,m})(b_2^{1,n}b_1^{n,m}-b_1^{1,n}b_2^{n,m})-(b_1^{n,m}-b_2^{n,m})(b_3^{1,n}b_1^{n,m}-b_1^{1,n}b_3^{n,m})}, \nonumber\\\nonumber
\lambda_n &=& \frac{(\overline{M}_3-\overline{M}_1)(b_2^{n,m}-b_1^{n,m})-(\overline{M}_2-\overline{M}_1)(b_3^{n,m}-b_1^{n,m})}
                 {(b_2^{n,m}-b_1^{n,m})(b_3^{1,n}-b_1^{1,n})-(b_3^{n,m}-b_1^{n,m})(b_2^{1,n}-b_1^{1,n})}, \nonumber\\\nonumber
\lambda_m &=& \frac{(\overline{M}_3-\overline{M}_1)(b_2^{1,n}-b_1^{1,n})-(\overline{M}_2-\overline{M}_1)(b_3^{1,n}-b_1^{1,n})}
                 {(b_3^{n,m}-b_1^{n,m})(b_2^{1,n}-b_1^{1,n})-(b_2^{n,m}-b_1^{n,m})(b_3^{1,n}-b_1^{1,n})},
\end{eqnarray}
where all the identities are interpreted as functional equalites 
on the set
$E$.
It is clear 
from~\eqref{eq:Symbol}
that the denominators in the above expressions satisfy
the inequalities
$$(b_1^{n,m}-b_3^{n,m})(b_2^{1,n}b_1^{n,m}-b_1^{1,n}b_2^{n,m})-(b_1^{n,m}-b_2^{n,m})(b_3^{1,n}b_1^{n,m}-b_1^{1,n}b_3^{n,m})<0,$$
$$(b_2^{n,m}-b_1^{n,m})(b_3^{1,n}-b_1^{1,n})-(b_3^{n,m}-b_1^{n,m})(b_2^{1,n}-b_1^{1,n})
<0,$$
for suffciently large
$m$
(e.g. 
$m\geq10$).
This is because the term 
$b_3^{n,m}$
dominates both expressions and has a negative coefficient in front of it.
We therefore find that, if the functions 
$\lambda_1,\lambda_n, \lambda_m$
are to be positive,
the following inequalities must be satisfied
%Given the above inequalities, in order for  to be non-negative we need
%\begin{eqnarray}
%(\overline{M}_3(x)b_1^{n,m}-\overline{M}_1(x)b_3^{n,m})(b_2^{1,n}b_1^{n,m}-b_1^{1,n}b_2^{n,m})-(\overline{M}_2(x)b_1^{n,m}-\overline{M}_1(x)b_2^{n,m})(b_3^{1,n}b_1^{n,m}-b_1^{1,n}b_3^{n,m})
%&<&0,\nonumber\\\nonumber
%(\overline{M}_3(x)-\overline{M}_1(x))(b_2^{n,m}-b_1^{n,m})-(\overline{M}_2(x)-\overline{M}_1(x))(b_3^{n,m}-b_1^{n,m})&<&0, \nonumber\\\nonumber
%(\overline{M}_3(x)-\overline{M}_1(x))(b_2^{1,n}-b_1^{1,n})-(\overline{M}_2(x)-\overline{M}_1(x))(b_3^{1,n}-b_1^{1,n})&>&0, \nonumber\\\nonumber
%\end{eqnarray}
%which imply
\begin{eqnarray}
\label{eq:ieq_lambda_1}
0 & < & \alpha^2 M_1(x)+\alpha M_2(x)\frac{b_3^{1,n}b_1^{n,m}-b_1^{1,n}b_3^{n,m}}{b_3^{n,m}b_2^{1,n}-b_3^{1,n}b_2^{n,m}}
-M_3(x)\frac{b_2^{1,n}b_1^{n,m}-b_1^{1,n}b_2^{n,m}}{b_3^{n,m}b_2^{1,n}-b_3^{1,n}b_2^{n,m}},\\
\label{eq:ieq_lambda_n}
0 & > & \alpha^2M_1(x)-\alpha M_2(x)\frac{b_3^{n,m}-b_1^{n,m}}{b_3^{n,m}-b_2^{n,m}}+M_3(x)\frac{b_2^{n,m}-b_1^{n,m}}{b_3^{n,m}-b_2^{n,m}},\\
\label{eq:ieq_lambda_m}
0 & < & \alpha^2M_1(x)-\alpha M_2(x)\frac{b_3^{1,n}-b_1^{1,n}}{b_3^{1,n}-b_2^{1,n}}+M_3(x)\frac{b_2^{1,n}-b_1^{1,n}}{b_3^{1,n}-b_2^{1,n}},
\end{eqnarray}
for every
$x\in E$.
These inequalities specify quadratic conditions
on the spacing
$\alpha$
(of the state-space of the process
$I$)
which have to be satisfied 
on the entire set
$E$.
%where $M_1(a),M_2(a),M_3(a)>0$ by definition and we can also find
%$$b_1^{n,m}(b_3^{1,n}b_2^{n,m}-b_3^{n,m}b_2^{1,n}))<0,\>\>\>b_3^{n,m}-b_2^{n,m}>0,\>\>\>b_3^{1,n}-b_2^{1,n}>0.$$
%Recall the solutions of the quadratic equation $A\alpha^2+B\alpha+C=0$ with $A,B,C\in\RR$
%are
%$$\alpha^\pm = {-B\pm\sqrt{B^2-4AC}\over 2A},$$
%and let $\overline{\alpha}:=\mathrm{max}(\alpha^+,\alpha^-)$,
%$\underline{\alpha}:=\mathrm{min}(\alpha^+,\alpha^-)$.
%We are now going to find the range of $\alpha$ which will give us 
%$\lambda_1,\lambda_n, \lambda_m\geq 0$. In other words, a range of $\alpha$
%which will satisfy the inequalities (\ref{eq:ieq_lambda_1}),
%(\ref{eq:ieq_lambda_n}) and (\ref{eq:ieq_lambda_m}) simultaneously for all
%$a\in\Omega$.

Note that inequality~\eqref{eq:ieq_lambda_1} is always satisfied if the corresponding
discriminant is negative.
%(i.e. $B^2<4AC$) because the coefficient of $\alpha^2$ is negative. 
Alternatively
if the discriminant is non-negative, % (i.e. $B^2\geq4AC$)
then the real zeros of the corresponding parabola,
denoted by
$\underline{\alpha}(x),\overline{\alpha}(x)$
and without loss of generality assumed to satisfy
$\underline{\alpha}(x)\leq \overline{\alpha}(x)$,
exist and 
the conditions 
$$\alpha < \underline{\alpha}(x)\quad\text{or}\quad\alpha > \overline{\alpha}(x)\quad\forall x\in E$$
must hold.
% for inequality (\ref{eq:ieq_lambda_1}) to be satisfied. 
Similar analysis
can be applied to inequality~\eqref{eq:ieq_lambda_m}. 
Inequality~\eqref{eq:ieq_lambda_n}
will always be violated if the discriminant is negative. 
This implies the following condition
\begin{equation}
\label{eq:max_Tx}
\frac{(b_3^{n,m}-b_1^{n,m})^2}{(b_3^{n,m}-b_2^{n,m})(b_2^{n,m}-b_1^{n,m})}\geq \frac{4M_1(x)M_3(x)}{M_2(x)^2}\quad\forall x\in E,
\end{equation}
which has to hold
regardless of the choice of the spacing
$\alpha$.
%Expression (\ref{eq:max_Tx}) 
%imposes an upper bound on 
%$M_3(x)$ for given 
%$n, m, M_1(x)$
%and $M_2(x)$. 
Even if condition~\eqref{eq:max_Tx} is satisfied
%If its discriminant is non-negative (i.e. inequality ()
%is satisfied) then 
we need to enforce the inequalities
$$\underline{\alpha}(x) < \alpha < \overline{\alpha}(x)\quad\forall x\in E.$$
%for all $a\in\Omega$ in order to satisfy inequality (\ref{eq:ieq_lambda_n}).
In Table~\ref{t:conditions}
we 
summarise the conditions that need to hold for
$\lambda_1(x),\lambda_n(x),\lambda_m(x)$ 
to be positive
for any fixed element 
$x\in E$.

\begin{table}[ht]
\begin{center}
\begin{tabular}{|c|c|c|c|c|} 
\hline 
& Discriminant $\geq0$   & Restriction on $\alpha$ \\
\hline
$\lambda_1(x)>0$ & true    & $\alpha < \underline{\alpha}(x)$ or $\alpha > \overline{\alpha}(x)$\\
%$\lambda_1(x)$ 
& false   & none\\
\hline
$\lambda_n(x)>0$ & true   
%${(M_2(x)(b_3^{n,m}-b_1^{n,m}))^2\over 4M_1(x)(b_3^{n,m}-b_2^{n,m})(b_2^{n,m}-b_1^{n,m})}\geq M_3(x)$
& $\underline{\alpha}(x) < \alpha < \overline{\alpha}(x)$\\
%$\lambda_n(x)$ 
& false  & $\lambda_n(x)$
cannot be positive \\
\hline
$\lambda_m(x)>0$ & true    & $\alpha < \underline{\alpha}(x)$ or $\alpha > \overline{\alpha}(x)$\\
%$\lambda_m(x)$ 
& false  & none \\
%& Discriminant $\geq0$  & Restriction on $M_3(x)$ & Restriction on $\alpha$ \\
%\hline
%$\lambda_1(x)$ & true   & none & $\alpha < \underline{\alpha}(x)$ or $\alpha > \overline{\alpha}(x)$\\
%$\lambda_1(x)$ & false  & none & none\\
%\hline
%$\lambda_n(x)$ & true   
%& condition~\eqref{eq:max_Tx} holds
%%${(M_2(x)(b_3^{n,m}-b_1^{n,m}))^2\over 4M_1(x)(b_3^{n,m}-b_2^{n,m})(b_2^{n,m}-b_1^{n,m})}\geq M_3(x)$
%& $\underline{\alpha}(x) < \alpha < \overline{\alpha}(x)$\\
%$\lambda_n(x)$ & false  & $\lambda_n(x)$ cannot be positive & $\lambda_n(x)$
%cannot be positive \\
%\hline
%$\lambda_m(x)$ & true   & none & $\alpha < \underline{\alpha}(x)$ or $\alpha > \overline{\alpha}(x)$\\
%$\lambda_m(x)$ & false  & none & none \\
\hline
\end{tabular}
\end{center}
\vspace{2mm}
\caption{\footnotesize{Conditions on 
the discriminant and the real roots
$\underline{\alpha}(x), \overline{\alpha}(x)$,
assumed to satisfy the relation
$\underline{\alpha}(x)\leq \overline{\alpha}(x)$,
in this table
refer to the parabolas that arise in inequalities~\eqref{eq:ieq_lambda_1},
\eqref{eq:ieq_lambda_n} and~\eqref{eq:ieq_lambda_m}.
These inequalities 
are equivalent to
the conditions
%in turn correspond to
$\lambda_1(x)>0$,
$\lambda_n(x)>0$
and
$\lambda_m(x)>0$
respectively.  }}
\label{t:conditions}
\end{table}
Having chosen 
the spacing
$\alpha$
according to the conditions in Table~\ref{t:conditions},
we can use the formulae above to compute 
functions
$\lambda_1, \lambda_n$
and
$\lambda_m$.
The conditional 
generator of the process
$I$, 
defined in~\eqref{eq:CondGenOfI},
now takes the form
%conditional on the underlying process being at the level $X(a)$ is now of the form
\begin{eqnarray*}
\L^I(x:c,d) := 
\left\{
\begin{array}{ll}
 \lambda_1(x) & \mathrm{if}\>\>d=(c+1)\!\!\!\mod (2C+1);\\
 \lambda_n(x) & \mathrm{if}\>\>d=(c+s)\!\!\!\mod (2C+1),\> s\in\{2,...,n\};\\               
 \lambda_m(x) & \mathrm{if}\>\>d=(c+s)\!\!\!\mod (2C+1),\> s\in\{n+1,...,m\};
%-\lambda_1(a)-(n-1)\lambda_n(a)  & \\ -(m-n)\lambda_m(a)&\mathrm{if}\>\>d=c,
\end{array}
\right. 
\end{eqnarray*}
with diagonal elements given by 
$-\lambda_1(x)-(n-1)\lambda_n(x)-(m-n)\lambda_m(x)$
and all other entries  equal to zero.
%and the expression (\ref{eq:generator_Ik_new_kth_moment})
%can be written explicitly as
%\begin{eqnarray}\label{eq:generator_Ik_third_moment}
%\L_k(a,b) 
%&:=& \L(a,b) + \delta_{a,b}
%\left(
%\lambda_0(a)
%+
%\lambda_1(a)e^{-\i p_k}
%+
%\lambda_n(a)\sum_{i=2}^n e^{-\i p_k i}
%+
%\lambda_m(a) \sum_{i=n+1}^m e^{-\i p_k i}
%\right),
%\nonumber\\
%\end{eqnarray}
In Section~\ref{sec:numerical}
we are going to implement the algorithm described here
for the variance gamma model and the subordinated CEV process.
%The values of 
%$n$
%and
%$m$
%will be approximately
%5
%and
%30 respectively.

\section{Numerical results}
\label{sec:numerical}
In this section we will perform a  numerical study
of the approximations given in the Sections~\ref{sec:second_moments}
and~\ref{sec:third_moments}. 
Subsection~\ref{subsec:MCapp}
gives an explicit construction of the approximating Markov chain
$X$
and compares the vanilla option prices with the ones in the original
model 
$S$.
Subsections~\ref{subsec:VolDer} and~\ref{subsec:VolDerJumps}
compare the algorithm for volatility derivatives described in this paper with 
a Monte Carlo simulation.

\subsection{Markov chain approximation}
\label{subsec:MCapp}

Let 
$S$
%be the CEV process. In other words 
%$S$
%is 
be a Markov process that 
satisfies SDE~\eqref{eq:SDE}
with the volatility function 
$\sigma:\RR_+\to\RR_+$
given by 
$\sigma(s):=\sigma_0 s^{\beta-1}$
and the drift 
$\gamma$
equal to the risk-free rate
$r$
(i.e. 
$S$
is a CEV process).
We generate the state-space
$E$
the algorithm
in Appendix~\ref{sec:NU_State_Space} 
and define
find the generator matrix
$\L$
by solving the linear system 
in~\eqref{eq:GenEqu}.
%where
%$r$
%is the risk-free rate and the function 
%$\sigma(s)$
%is as defined above.
%Note that if the parameter
%$\beta=1$,
%then the described procedure yields a Markov chain
%approximation for the geometric Brownian motion, i.e.
%a Black-Scholes model.

If the process
$S$
is a jump-diffusion of the form described in Subsection~\ref{subsec:Jump_Diff}
(e.g. a variance gamma model or a CEV model subordinated by a gamma process),
we obtain the generator for the chain 
$X$
by applying formula~\eqref{eq:SubDiag}
to the generator defined in the previous paragraph,
where the function 
$\phi$
(in~\eqref{eq:SubDiag})
is given by~\eqref{bochner}.
More preciselly
if
$S$
is the subordinated 
CEV process, the drift 
$\gamma$
in~\eqref{eq:SDE}
is given by the formula in~\eqref{eq:DriftSub}.
If
$S$
is a variance gamma model
we subordinate the geometric Brownian motion 
which solves 
SDE~\eqref{eq:SDE} 
with the constant volatility function
$\sigma(s)=\sigma_0$
and the drift
$\gamma=\theta+\sigma_0^2/2$
where 
$\theta$
is the parameter in the vairance gamma model
(see~\cite{MCC}, equation~(1)).
The implementation in Matlab of this 
construction can be found in~\cite{LM_Code}.
Note that the state-space of the Markov
chain 
$X$
in the diffusion and the jump-diffusion cases is 
of the same form (i.e. given by the algorithm in Appendix~\ref{sec:NU_State_Space}).

The numerical accuracy of these approximations is illustrated in 
Tables~\ref{t:cev_ivol},
\ref{t:vg_ivol}
and~\ref{t:SubCEV_ivol}
where the vanilla option prices in the Markov chain model
$X$
are compared with the prices in the original model
$S$
for the CEV process, the variance gamma model and the subordinated 
CEV model
respectively.

%Follows \cite{AM}, the generator $\L$ for the Black-Scholes ($\beta=1$) and CEV process can be obtained by solving the following systems:
%\begin{eqnarray}
%  \sum_{b\in\Omega} \L(a,b) &=& 0, \\
%  \sum_{b\in\Omega} \L(a,b) (X(a)-X(b)) &=& r X(a), \\
%  \sum_{b\in\Omega} \L(a,b) (X(a)-X(b))^2 &=&  \sigma^2
%                   \left( {X(a)\over s}\right)^{2\beta-2} X(a)^2,
%\end{eqnarray} 
%for all $a\in\Omega$, where $r$ is the risk free rate, $\sigma$ is the volatility and $s$ is the spot price. The generator of the VG process can be obtained by applying the corresponding
%Bernstein function to the Black-Scholes generator, see \cite{AM}.

%Before pricing any products related to the realized variance, 
%one would need to examine the accuracy of the underlying model. 
%The following tables present the implied volatilities obtain from 
%exponentiate the underlying generator 
%$\L$ 
%and compare against the closed-form or quasi-closed-form formula:

\begin{table}[ht]
\begin{center}
\scalebox{0.8}{
\begin{tabular}{|c|ccc|ccc|}
\hline 
&  & Markov chain $X$ &  &  & CEV: closed-form &  \\
\hline
$K \backslash T$ & $0.5$ & $1$ & $2$ & $0.5$ & $1$ & $2$ \\
\hline
80 & 21.44\% & 21.42\% & 21.30\% & 21.54\% & 21.47\% & 21.34\% \\
90 & 20.55\% & 20.57\% & 20.46\% & 20.68\% & 20.62\% & 20.49\% \\
100 & 19.93\% & 19.90\% & 19.71\% & 19.94\% & 19.88\% & 19.75\% \\
110 & 19.37\% & 19.19\% & 19.11\% & 19.28\% & 19.22\% & 19.10\% \\
120 & 18.76\% & 18.66\% & 18.53\% & 18.69\% & 18.63\% & 18.52\% \\
%  &  & Markov chain $X$ &  &  & CEV: closed-form &  \\
%\hline 
%$K \backslash T$ & $0.5$ & $1$ & $2$ & $0.5$ & $1$ & $2$ \\
%\hline
%80 & 21.17\% & 21.16\% & 21.15\% & 21.16\% & 21.15\% & 21.15\% \\
%90 & 20.55\% & 20.54\% & 20.54\% & 20.54\% & 20.54\% & 20.55\% \\
%100 & 19.99\% & 19.99\% & 19.99\% & 20.01\% & 20.01\% & 20.02\% \\
%110 & 19.53\% & 19.51\% & 19.53\% & 19.54\% & 19.54\% & 19.55\% \\
%120 & 19.11\% & 19.10\% & 19.10\% & 19.12\% & 19.11\% & 19.12\% \\
  \hline
\end{tabular}
} % end scale box
\end{center}
\vspace{2mm}
\caption{\footnotesize{Implied volatility in the CEV model. 
The maturity 
$T$ 
varies from half a year to two years
and
the corresponding strikes
are of the form
$Ke^{rT}$,
where
$K$
takes values between 
80
and
120 
and
the
risk-free rate equals
$r=2\%$.
The CEV process 
$S$,
with the current spot value
$S_0=100$,
is given by~\eqref{eq:SDE}
with the local volatility function 
$\sigma$
equal to
$\sigma(s):=\sigma_0 s^{\beta-1}$
and the
drift
$\gamma=r$,
where 
the volatility parameters are
$\sigma_0=0.2$, $\beta=0.3$. 
The parameters for the non-uniform state-space 
of the chain
$X$
are
$N=70$ 
and
$l=1, s=100, u=700, g_l=50, g_u=50$
(see Appendix~\ref{sec:NU_State_Space}
for the definition of these parameters)
and the generator of 
$X$
is specified by  system~\eqref{eq:GenEqu}.
The pricing in the Markov chain model is 
done using~\eqref{eq:MarkovChainSemigroup}
and  in the CEV model using
a closed-form formula in~\cite{Hull},
pages 562-563.
The total computation time for all the 
option price in the table under the Markov chain model 
$X$
is less than one tenth of a second
on a standard PC with 1.6GHz Pentium-M processor and 1GB RAM.
}}
\label{t:cev_ivol}
\end{table}

\begin{table}[ht]
\begin{center}
\scalebox{0.8}{
\begin{tabular}{|c|ccc|ccc|}
\hline 
&  & Markov chain $X$ &  &  & VG: FFT &  \\
\hline
$K \backslash  T $ & $0.5$ & $1$ & $2$ & $0.5$ & $1$ & $2$ \\
\hline
80 & 20.43\% & 20.07\% & 19.98\% & 20.44\% & 20.09\% & 20.00\% \\
90 & 19.91\% & 19.89\% & 19.93\% & 19.95\% & 19.94\% & 19.96\% \\
100 & 19.69\% & 19.84\% & 19.92\% & 19.75\% & 19.87\% & 19.94\% \\
110 & 19.85\% & 19.89\% & 19.93\% & 19.82\% & 19.88\% & 19.93\% \\
120 & 20.16\% & 19.92\% & 19.94\% & 20.08\% & 19.93\% & 19.94\% \\
%  &  & Markov chain $X$ &  &  & VG: FFT &  \\
%\hline 
%$K \backslash  T $ & $0.5$ & $1$ & $2$ & $0.5$ & $1$ & $2$ \\
%\hline
%80 &	20.50\%	& 20.08\% & 20.02\% & 20.44\% &	20.09\% & 20.00\% \\
%90 &    19.90\% & 19.91\% & 19.95\% & 19.95\% &	19.94\% & 19.96\% \\
%100 &	19.69\%	& 19.85\% & 19.92\% & 19.75\% &	19.87\% & 19.94\% \\
%110 &	19.84\%	& 19.87\% & 19.93\% & 19.82\% &	19.88\% & 19.93\% \\
%120 &	20.06\%	& 19.94\% & 19.93\% & 20.08\% &	19.93\% & 19.94\% \\
  \hline
\end{tabular}
} % end scale box
\end{center}
\vspace{2mm}
\caption{\footnotesize{Implied volatility in the variance gamma model. 
The strikes %$K$
and maturities % $T$
are as in Table~\ref{t:cev_ivol}.
The process 
$S$,
with the current spot value
$S_0=100$,
is obtained 
by subordinating diffusion~\eqref{eq:SDE}
with the constant volatility function
$\sigma(s)=\sigma_0$
and 
the drift
equal to 
$\gamma=\theta+\sigma_0^2/2$,
where
$\theta$
is given in~\cite{MCC}, equation~(1).
The Bernstein function of the gamma 
subordinator
is given in~\eqref{bochner}.
%and the subordinated process has to be mean-corrected
%as specified in Formula~(22) of~\cite{MCC}.
The
risk-free rate is assumed to be
$r=2\%$,
the  
diffusion
parameters
take values 
$\sigma_0=0.2, \theta=-0.04,$ 
and
the jump parameters in~\eqref{bochner}
equal
$\mu=1, \nu=0.05$. 
The parameters for the state-space 
of the chain
$X$
are
$N=70$
and
$l=1, s=100, u=700,$ 
$g_l=30, g_u=30$
(see Appendix~\ref{sec:NU_State_Space}
for the definition of these parameters).
%$l=1, s=100, u=500, N=70, g_l=200, g_u=25$
%(see Appendix~\ref{sec:NU_State_Space}
%for the definition of these parameters)
%and the generator matrix
%$\L$
%of 
%$X$
%is specified by system~\eqref{eq:GenEqu}.
The total computation time for all the 
option price in the table under the Markov chain model 
is less than one tenth of a second.
The Fourier inversion is performed using the algorithm in~\cite{CarrMadan}
and takes approximately
the same amount of time.
All computations are performed 
on the same hardware as in Table~\ref{t:cev_ivol}.
}}
\label{t:vg_ivol}
\end{table}

\begin{table}[ht]
\begin{center}
\scalebox{0.8}{
\begin{tabular}{|c|ccc|ccc|}
\hline 
&  & Markov chain $X$ &   &  & CEV with jumps: MC &  \\
\hline
$K \backslash  T $ & $0.5$ & $1$ & $2$ & $0.5$ & $1$ & $2$ \\
\hline
80 & 20.82\% & 20.57\% & 20.49\% & 20.92\% & 20.66\% & 20.41\% \\
90 & 20.08\% & 20.10\% & 20.11\% & 20.16\% & 20.12\% & 20.19\% \\
100 & 19.74\% & 19.83\% & 19.82\% & 19.75\% & 19.81\% & 19.78\% \\
110 & 19.66\% & 19.61\% & 19.56\% & 19.64\% & 19.58\% & 19.53\% \\
120 & 19.72\% & 19.48\% & 19.32\% & 19.75\% & 19.37\% & 19.39\% \\
%  &  & Markov chain $X$ &   &  & CEV with jumps: MC &  \\ 
%\hline 
%$K \backslash  T $ & $0.5$ & $1$ & $2$ & $0.5$ & $1$ & $2$ \\
%\hline
%80 &	20.94\% & 20.66\% &	20.63\% &	21.13\% &	20.62\% &	20.53\% \\
%90 &	20.09\% & 20.16\% &	20.23\% &	20.21\% &	20.17\% &	20.22\% \\
%100 &	19.65\% & 19.82\% &	19.91\% &	19.69\% &	19.77\% &	19.93\% \\
%110 &	19.55\% & 19.59\% &	19.64\% &	19.74\% &	19.62\% &	19.73\% \\
%120 &	19.77\% & 19.49\% &	19.44\% &	19.79\% &	19.43\% &	19.49\% \\
  \hline
\end{tabular}
} % end scale box
\end{center}
\vspace{2mm}
\caption{
\footnotesize{Implied volatility in the CEV model subordinated 
by a gamma process. 
The strikes %$K$
and maturities %$T$
are as in Table~\ref{t:cev_ivol}.
The process 
$S$,
with the current spot value
$S_0=100$,
is obtained 
by subordinating diffusion~\eqref{eq:SDE}
with the volatility function
$\sigma(s)=\sigma_0 s^{\beta-1}$
(where
$\sigma_0=0.2, \beta=0.7$)
and 
the drift
given by~\eqref{eq:DriftSub}
(where
the risk-free rate is 
$r=2\%$
and the jump-parameters in~\eqref{bochner}
equal
$\mu=1, \nu=0.05$). 
The parameters for the state-space 
of the chain
$X$
are
as in Table~\ref{t:vg_ivol}.
The total computation time for all the 
option price in the table under the Markov chain model 
is less than one tenth of a second.
The prices in the model 
$S$
were computed using a Monte Carlo algorithm that first generates
the paths of the gamma process
$(T_t)_{t\geq0}$
(using the algorithm  
in~\cite{GlassMC}, page 144)
and then, via an Euler scheme, generates paths of the process
$S$.
For the 
$T=2$
years
maturity,
$10^5$
paths were generated in 
$200$ 
seconds.
All computations are performed 
on the same hardware as in Table~\ref{t:cev_ivol}.
}}
\label{t:SubCEV_ivol}
%{t:black_scholes_ivol}
\end{table}

It is clear from 
Tables~\ref{t:cev_ivol},
\ref{t:vg_ivol}
and~\ref{t:SubCEV_ivol}
that the continuous-time Markov chain 
$X$
%briefly described in Sections~\ref{sec:second_moments}
%and~\ref{sec:third_moments},
approximates reasonably well
the Markov process
$S$
on the level of 
European option prices. 
The pricing in the Markov chain model is done 
by matrix exponentiation. 
The transition semigroup of the 
chain 
$X$
is of the form
\begin{eqnarray}
\label{eq:MarkovChainSemigroup}
\PP(X_t=y\lvert X_0=x)=e_x'\exp(t\L)e_y,
\quad\text{where}\quad
x,y\in E,
\end{eqnarray}
$e_x,e_y$
are the corresponding vectors of the standard basis
of
$\RR^N$
and
$'$
denotes transposition.  
For more details on this pricing algorithm  see~\cite{AM}.
The implied volatilities in the CEV, the variance gamma and
the subordinated CEV model were obtained by a closed-form 
formula, a fast Fourier transform inversion algorithm and a
Monte Carlo algorithm respectively.
As mentioned earlier 
the quality of this approximation can be improved considerably, 
without increasing the size of the set
$E$,
by matching more than the first two instantaneous 
moments of the process
$S$.

%The closed-form 
%solution for European option in the CEV model can be found in p.562-563
%of \cite{Hull} and the European option in the VG model are priced using the
%FFT method suggested by \cite{CarrMadan}.

\subsection{Volatility derivatives -- the continuous case}
\label{subsec:VolDer}

The next task is to construct the %approximate volatility 
process
$I$
defined in Section~\ref{sec:k_moments},
obtain its law at a maturity
$T$
using Theorem~\ref{thm:LiftKer}
and compare it to the law of the random variable
$[\log(S)]_T$
%or
%$Q^{L,U}_T(S)$,
defined in~\eqref{eq:QuadVarS}
%and~\eqref{eq:CorrRealVar}
%respectively,
by pricing non-linear contracts.
%obtain the probability density functions of the realized
%variance using our methods. 

%We start with the diffusion case.
Let 
$S$
be the CEV processs with the parameter values as in the caption of
Table~\ref{t:cev_ivol}
and let
$X$
be the corresponding Markov chain,
which is also described uniquely in the same caption.
As described in 
Section~\ref{sec:second_moments}
in this case we use
$k=2$
(i.e.
the process
$I$
matches the first and the second instantaneous
conditional moments of the process
$[\log(X)]_T$
%$Q^{L,U}(X)$
defined in~\eqref{eq:QVMarkovChain})
and hence 
define the state-dependent 
intensities
in the conditional generator of 
$\L^I$ 
by~\eqref{eq:l1Diff}
and~\eqref{eq:lnDiff}. 
We still need to determine the values
of 
the spacing
$\alpha$,
the size 
$(2C+1)$
of the state-space 
of 
$I$
and
the largest possible jump-size
$\alpha n$
of the process
$I$
at any given time.

\begin{figure}[ht]
\begin{center}
\subfloat[\footnotesize{The probability distribution function for the spot
                 price 
		 $X_T$,
		 with the maturity
		 $T$
		 equal to 0.5, 1 and 2 years, 
		 where
	         $X$
		 is the Markov 
		 chain  used to approximate the CEV process
		 $S$. 
		 For a precise description of the process
		 $X$
		 see Subsection~\ref{subsec:MCapp}.
		 All relevant parameter values
		 are given in the caption of Table~\ref{t:cev_ivol}.}]{\includegraphics[width=8cm]{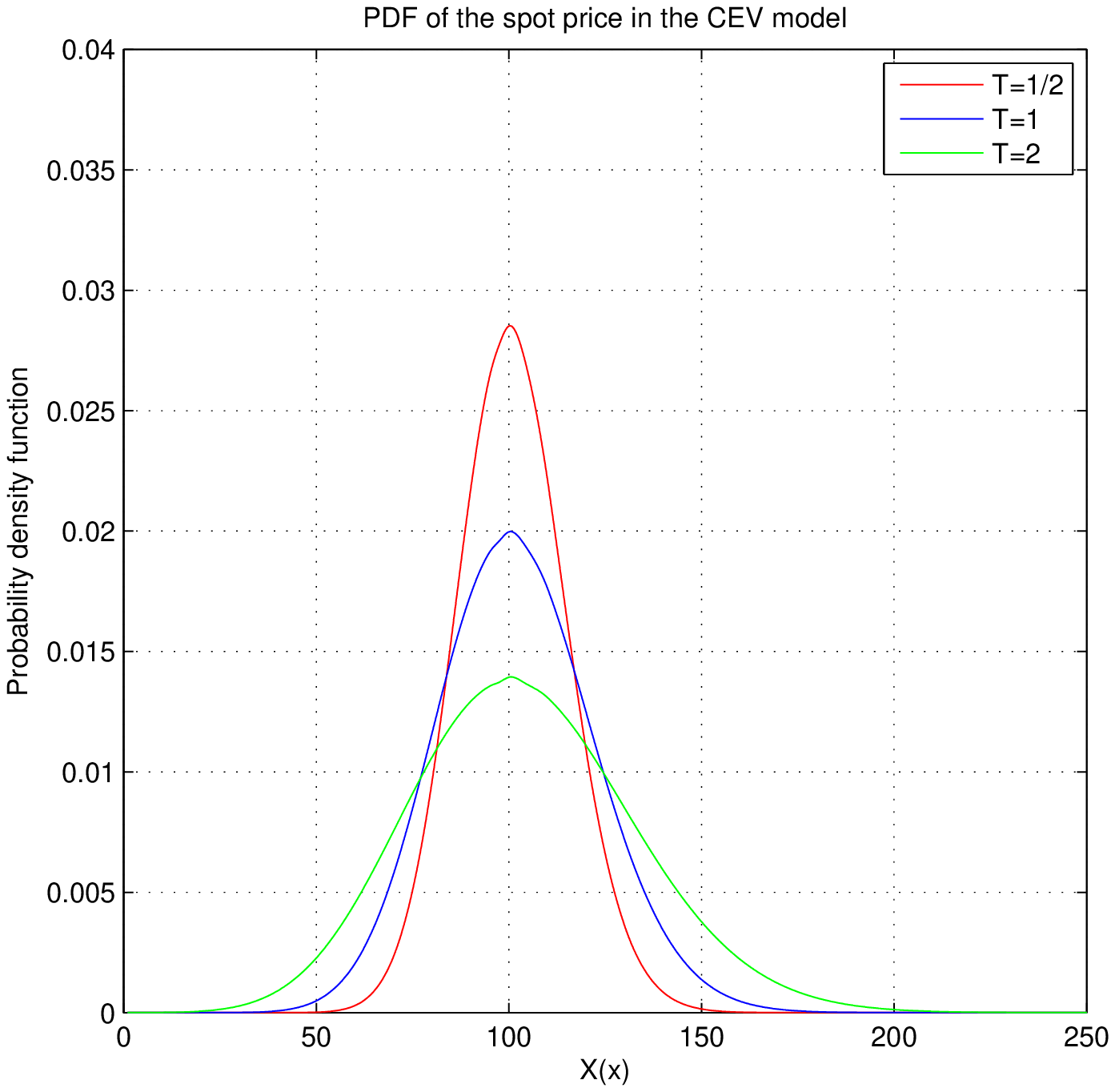}\label{fig:cev_pdf}}
\hspace{4mm}\subfloat[\footnotesize{The function $x\mapsto M_2(x)/M_1(x)$, where $x\in E$, 
                 in the CEV model. The minimum of this function, which equals
		 $0.000563$,
		 determines the value of the spacing $\alpha$ 
		 by the second inequality in~\eqref{eq:FindAlphaDiff}.
		 The maximum of the ratio, which is $0.019$, 
		 determines the largest jump-size multiple
		 $n$ 
		 by the first inequality 
                 in~\eqref{eq:FindAlphaDiff}.
		 All relevant parameter values for the CEV model 
		 and the accompanying chain
		 $X$
		 are given in the caption of Table~\ref{t:cev_ivol}.}]{\includegraphics[width=8cm]{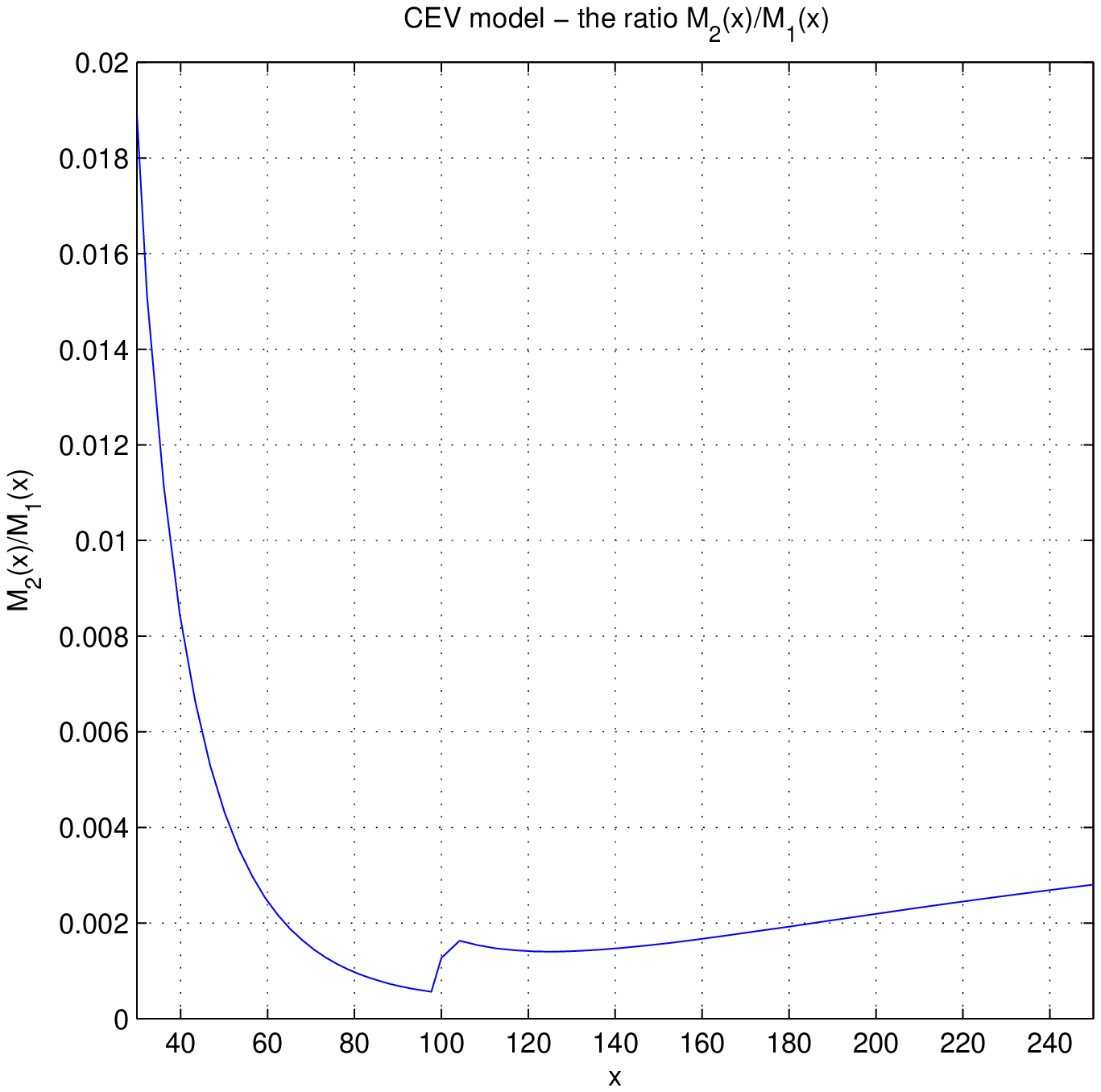}\label{fig:cev_VQ_ratio}}\\
\subfloat[\footnotesize{The empirical probability distribution of the realized
    variance 
    $[\log(S)]_T$
    of the CEV model
    $S$,
    based on the Monte Carlo simulation described in Subsection~\ref{subsec:VolDer},
    and the distribution of the random variable 
    $I_T$,
    obtained from Theorem~\ref{thm:LiftKer},
    for 
    $T\in\{0.5,1,2\}$.
    For details on the definition of
    $I_T$
    see Sections~\ref{sec:k_moments} and~\ref{sec:second_moments}.
    Note that the computational time required to obtain the law of
    $I_T$
    is independent of 
    $T$
    (see caption of Table~\ref{t:cev_varpayoff}).}]{\includegraphics[width=10cm]{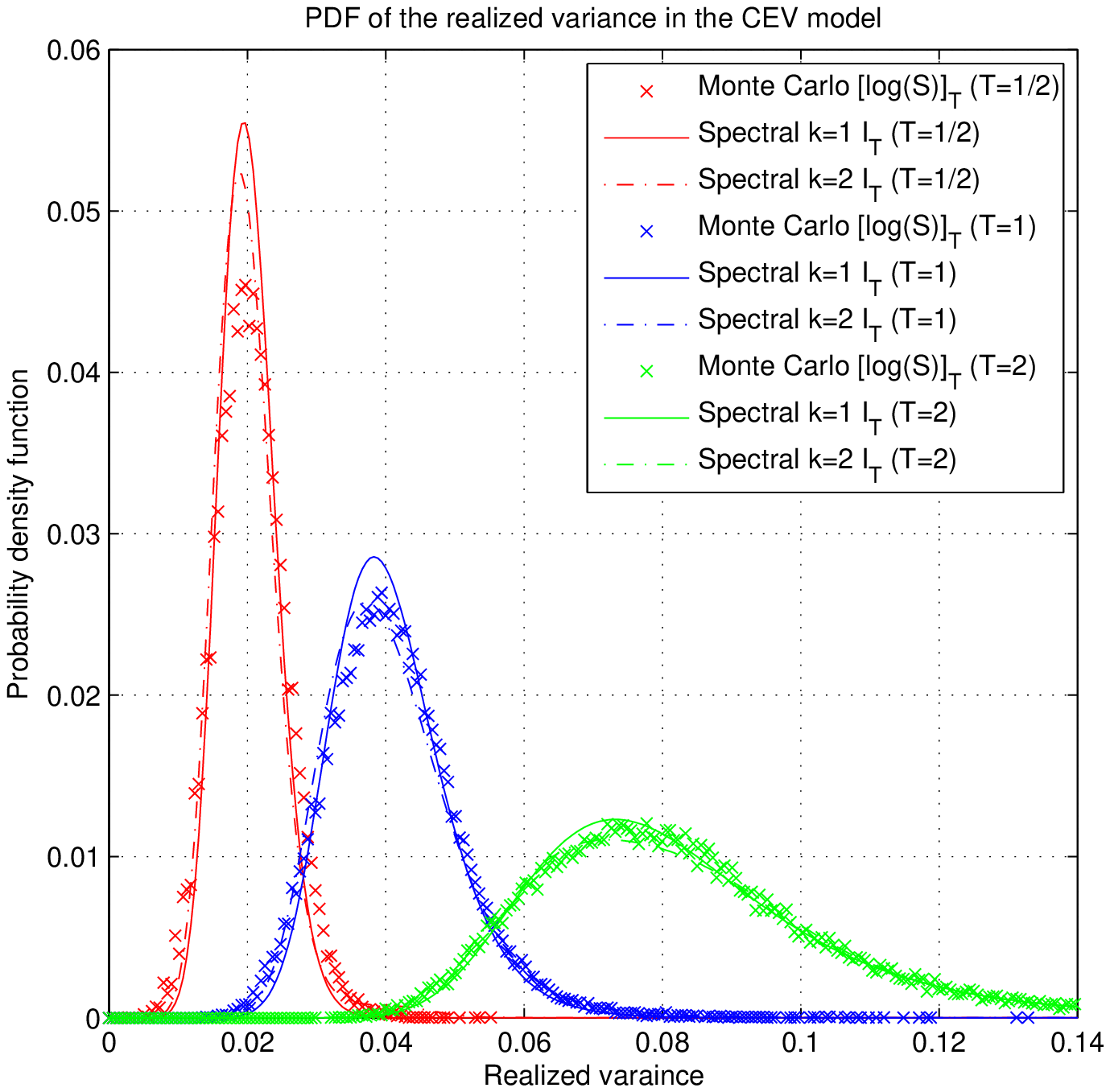}\label{fig:cev_var_pdf}}
\end{center}
\caption{\footnotesize{CEV model}
%Figure~\ref{fig:cev_var_pdf} contains the empirical probability distribution of the realised
%    variance 
%    $[\log(S)]_T$
%    of the CEV model
%    $S$,
%    based on the Monte Carlo simulation described in Subsection~\ref{subsec:VolDer},
%    and the distribution of the random variable 
%    $I_T$,
%    obtained from Theorem~\ref{thm:LiftKer},
%    for maturity 
%    $T\in\{0.5,1,2\}$.
%    For details on the definition of
%    $I_T$
%    see Sections~\ref{sec:k_moments} and~\ref{sec:second_moments}.
%    Note that the computational time required to obtain the law of
%    $I_T$
%%    is independent of 
%    $T$
%    and in this case equals 23 seconds
%    (see caption of Table~\ref{t:cev_varpayoff} for more details).}
}
\end{figure}

The necessary and sufficien condition on parameters
$\alpha$
and
$n$
is given by~\eqref{eq:FindAlphaDiff}.
Figure~\ref{fig:cev_VQ_ratio} contains the graph of the ratio in question
$x\mapsto M_2(x)/M_1(x)$,
$x\in E$,
for the CEV model.
The minimum of the
ratio is 
$0.000563$,  
which can be used to define the value of 
$\alpha$. 
The largest
value of the ratio is approximately 
$0.019$ 
and hence
$n=50$ 
satisfies the first inequality in~\eqref{eq:FindAlphaDiff}.
%(note that $\alpha b_2^{1,35}/b_1^{1,35}=0.006\cdot23.7=0.0142$).

An important observation here is that Figure~\ref{fig:cev_VQ_ratio}
only displays the values of the ratio 
$ M_2(x)/M_1(x)$
for 
$x$
in 
$E\cap[20,250]$.
The choices of 
$\alpha$
and
$n$
made above therefore satisfy condition~\eqref{eq:FindAlphaDiff}
only in this range
(recall that in this case we have
$x_0=1$ and 
$x_{69}=700$).
However this apparent violation of
the condition in~\eqref{eq:FindAlphaDiff}
plays no role 
because the probability for the underlying 
process
$X$ 
to get below 
$20$ 
or above 
$250$ 
in 2 years time is less than $10^{-6}$
(see Figure~\ref{fig:cev_pdf}). 
This intuitive statement is supproted by the quality of the
approximation of the empirical distribution of
$[\log(S)]_T$
by the distribution of
$I_T$
(see Figure~\ref{fig:cev_var_pdf}
and Table~\ref{t:cev_varpayoff}).
%Therefore the violation of condition ~\eqref{eq:FindAlphaDiff}
%The parameters
%$n$ and $\alpha$ are not chosen to match the second moment of the realized variance for $X(a)<20$ and $X(a)>250$ explicitly. Put it differently, it means
%that conditions on the first moment is $M_1(a)$, the second moment of the approximating
%Markov chain $I_t$ is
%$$\bar M_2(a) = \mathrm{min}\left(\mathrm{max}\left(M_2(a), \alpha\right)
%           , \alpha\left(\frac{\sum_{i=2}^n i^2}{\sum_{i=2}^n i}\right)\right).$$
%Our numerical experiments suggest this truncation does not have any significant
%impact on the price of the payoff depends on the realized variance with maturity less than or equal to 2 years.

We now need to choose the size 
$(2C+1)$ 
of the state-space for the process
$I$.
The integer
$C$
is determined by the
longest maturity 
that we are interested in,
which in our case is
2 years. 
This is because we are using Theorem~\ref{thm:LiftKer}
to find the joint law of the random variable
$(X_T,I_T)$
and must make sure that the process 
$I$
does not complete the full circle during the time interval 
of length
$T$
(recall that the pricing algorithm based on
Theorem~\ref{thm:LiftKer}
makes the assumption that the process
$I$
is on a circle).
In other words we 
have to choose  
$C$
so that the chain
$X$
accumulates much less than
$2C\alpha$
of realized variance.
In the example considered here it is sufficient to take
$C=220$,
which makes the state-space
$\{0,\alpha,\ldots,\alpha 2C\}$,
defined in the paragraph following~\eqref{eq:CondMomentSigma},
a uniform lattice in the interval
between
$0$
and
$440\cdot 0.00056=0.246$.
%it is sufficient to take $241 = 2C + 1$ uniformly distributed points for
%the grid $\Psi$. 
Since the spacing 
$\alpha$ 
does not change with maturity,
all that is needed to obtain the joint 
probability distribution of
$(X_T,I_T)$
for all maturities 
$T\in\{0.5, 1,2\}$ 
is to 
diagonalize numerically the complex matrices
$\L_j$,
$j=0,...,2C$,
in~\eqref{eq:joint_pdf} 
only once.
The distribution of 
$I_T$,
obtained as a marginal of the random vector
$(X_T,I_T)$,
is plotted in Figure~\ref{fig:cev_var_pdf}.
Note that the computational time required to 
obtain the law of
$I_T$
is therefore independent of maturity 
$T$.

\begin{table}[ht]
\begin{center}
\scalebox{0.7}{
\begin{tabular}{|c|c|ccc||ccc|ccc|}
\hline 
\textbf{CEV}  &  $k$ moments & & Spectral: $I_T$ &  &  & MC:
$[\log(S)]_T$ &  \\
\hline
$\mathrm{derivative} \backslash T $ & & $0.5$ & $1$ & $2$ & $0.5$ & $1$ & $2$ \\
\hline
var swap & 1 & 20.07\% & 20.19\% & 20.43\% & 20.09\% & 20.20\% & 20.42\%\\
$\sqrt{\EE[\Sigma_T/T]}$ & 2 & 20.07\%  & 20.19\% & 20.42\% & (0.051\%) & (0.051\%) & (0.052\%)\\
\hline
vol swap & 1 & 19.97\% & 20.08\% & 20.25\% & 19.92\% & 20.06\% & 20.22\%\\
$\EE\left[\sqrt{\Sigma_T/T}\right]$  & 2 & 19.92\% & 20.05\% & 20.22\% & (0.006\%) & (0.007\%) & (0.009\%)\\
\hline
call option & 1 & 1.46\% & 1.47\% & 1.51\% & 1.46\% & 1.48\% & 1.53\%\\
$\theta=80\%$ & 2 & 1.46\% & 1.47\% & 1.52\% &  (0.003\%) & (0.003\%) & (0.005\%)\\
\hline
call option & 1 & 0.33\% & 0.33\% & 0.43\% & 0.39\% &0.38\% & 0.45\%\\
$\theta=100\%$ & 2 & 0.38\% & 0.38\% &  0.45\% & (0.002\%) & (0.002\%) & (0.004\%)\\
\hline
call option & 1 & 0.01\% & 0.02\% & 0.07\% & 0.05\% & 0.03\% & 0.08\%\\
$\theta=120\%$ & 2 & 0.06\% & 0.04\% &  0.08\% & (0.001\%) & (0.001\%) & (0.003\%)\\
\hline
\hline
Time & & &  15s & &50s &100s &200s\\ 
%\textbf{CEV}  &  $k$ moments & & Spectral: $I_T$ &  &  & MC: $[\log(S)]_T$ &  \\
%\hline 
%$\mathrm{derivative} \backslash T $ & & $0.5$ & $1$ & $2$ & $0.5$ & $1$ & $2$ \\
%\hline
%var swap & 1 & 20.04\%	& 20.10\% & 20.21\% &	20.05\% & 20.11\% & 20.23\%\\
%$\sqrt{\EE[\Sigma_T/T]}$ & 2 & 20.04\%	& 20.10\% &	20.20\% & (0.072\%) & (0.072\%) & (0.072\%)\\
%\hline
%vol swap & 1 & 19.99\% & 20.08\% & 20.17\% & 19.90\% & 20.00\% & 20.11\%\\
%$\EE\left[\sqrt{\Sigma_T/T}\right]$  & 2 & 19.90\% & 20.00\% & 20.09\% & (0.011\%) & (0.009\%) & (0.010\%)\\
%\hline
%call option & 1 & 1.41\% & 1.42\% & 1.41\% & 1.47\% & 1.46\% & 1.48\%\\
%$\theta=80\%$ & 2 & 1.46\% & 1.46\% & 1.47\% &	(0.004\%) & (0.004\%) &	(0.004\%)\\
%\hline
%call option & 1 & 0.25\% & 0.22\% & 0.22\% & 0.38\% &0.32\% & 0.34\%\\
%$\theta=100\%$ & 2 & 0.37\% & 0.32\% &	0.33\% & (0.003\%) & (0.002\%) & (0.003\%)\\
%\hline
%call option & 1 & 0.00\% & 0.00\% & 0.01\% & 0.02\% & 0.01\% & 0.03\%\\
%$\theta=120\%$ & 2 & 0.03\% & 0.02\% &	0.02\% & (0.001\%) & (0.001\%) & (0.001\%)\\
%  \hline
%  \hline
%Time & & &  23s & &25s &50s &100s\\
\hline
\end{tabular}
} % end scale box
\end{center}
\vspace{2mm}
\caption{\footnotesize{The prices of volatility derivatives  in the CEV
model
$S$.  
The parameter values for the process 
$S$
and the chain
$X$
are given in the caption of Table~\ref{t:cev_ivol}.
The parameters for the process
$I$ 
are 
$\alpha=0.00056$, $C=220$ 
for
$k\in\{1,2\}$
and
$n=50$
when
$k=2$
(recall from Section~\ref{sec:second_moments}
that the parameter
$n$
controls
the jumps of 
$I$
strictly larger than 
$\alpha$,
which are note present if
$k=1$).
The variable
$\Sigma_T$
denotes either 
$I_T$
or
$[\log(S)]_T$
and the call option price is
$\EE\left[(\Sigma_T/T-(\theta K_0)^2)^+ \right]$,
for 
$\theta\in\{80\%, 100\%, 120\%\}$, 
$K_0:= \sqrt{\EE[\Sigma_T/T]}$.
An Euler scheme with a time-increment of 
one day is used to generate
$10^5$
paths
of the CEV process 
$S$
and the sum in~\eqref{eq:QuadVarS} is used
to obtain the empirical distribution of
$[\log(S)]_T$
(see Figure~\ref{fig:cev_var_pdf})
and to 
evaluate the contingent claims 
in this table. 
The numbers in brackets are the standard errors in 
the Monte Carlo simulation.
The computational time for the pricing of volatility 
derivatives using our algorithm
is independent of the maturity
$T$.
All computations are performed on the same hardware
as in Table~\ref{t:cev_ivol}.  }}
\label{t:cev_varpayoff}
\end{table}

We now perform a numerical 
comparisons between our method for pricing volatility 
derivatives 
and  a pricing algorithm based on 
a Monte Carlo simulation of the CEV model 
$S$.
We generate 
$10^5$
paths
of the process
$S$
using an Euler scheme
and compute the empirical probability distribution of
the realized variance
$[\log(S)]_T$
%defined by the
%sum in~\eqref{eq:QuadVarS},
based on that sample
(see Figure~\ref{fig:cev_var_pdf}).
We also compute the variance swap, the volatility swap
and the call option prices
$\EE\left[(\Sigma_T/T-(\theta K_0)^2)^+ \right]$,
for 
$\theta\in\{80\%, 100\%, 120\%\}$, 
where
$K_0:= \sqrt{\EE[\Sigma_T/T]}$ 
and
$\Sigma_T$
denotes either 
$I_T$
or
$[\log(S)]_T$.
The prices and the computation times are documented in Table~\ref{t:cev_varpayoff}.
A cursory inspection 
of the prices of non-linear payoff functions
reveals that the method for
$k=2$
outperforms the algorithm proposed in~\cite{ALM},
which corresponds to 
$k=1$,
%in the case of non-linear payoff functions
without adding computational complexity 
since both algorithms require
finding the spectrum of 
$(2C+1)$
complex matrices in~\eqref{eq:generator_Ik_new_kth_moment}.
We will soon see that 
the discrepancy between the algorithm in~\cite{ALM}
and the one proposed in the current paper is 
amplified in the presence of jumps.
Note also that all three methods
($k=1,2$
and the Monte Carlo method)
agree in the case of linear payoffs.

%For the Monte
%Carlo method we generate paths for the underlying with
%maturities 0.5, 1 and 2 years. Using those
%paths we price variance swaps $\sqrt{{1\over T}\EE_0[\Sigma_T]}$, quoted
%in volatility terms, volatility swaps $\EE_0\left[\sqrt{{1\over T}\Sigma_T}\right]$
%and call options on the realized variance
%$\EE_0\left[\mathrm{max}\left\{ {1\over T}\Sigma_T-(\theta K_0)^2,0\right\} \right]$,
%where the strike $K_0$ equals to the current risk-neutral mean of the realized
%variance and the parameter $\theta$ equals $80\%, 100\%, 120\%$. 
%The results are given in 
%Table \ref{t:black_scholes_varpayoff}. 
%Similar analysis can be done for the CEV model 
%and the results are presented in Table \ref{t:cev_varpayoff}. 
%Notice that the computation times of our method are for all 
%maturities (0.5, 1 and 2 years) since we only need to diagonalized 
%the matrices $\L_k$ once as discussed previously.

\subsection{Volatility derivatives -- the discontinuous case}
\label{subsec:VolDerJumps}
In this subsection we will study numerically the behaviour of
the law of random variables
$[\log(S)]_T$
and
$Q^{L,U}_T(S)$,
defined in~\eqref{eq:QuadVarS}
and~\eqref{eq:CorrRealVar}
respectively,
where 
$S$
is a Markov process with jumps.
Let 
$S$
be a variance gamma or a subordinated CEV process with parameter values
given in the captions of Tables~\ref{t:vg_ivol} and \ref{t:SubCEV_ivol}
respectively.

Since 
$S$
has discontinuous trajectories
we will have to match 
$k=3$
instantaneous conditional moments 
when defining the process
$I$
in order to avoid large pricing errors
for non-linear payoffs
(see Tables~\ref{t:vg_varpayoff}
and~\ref{t:SubCEV_varpayoff} for the size of the errors
when
$k=2$). 
We firts define the Markov chain 
$X$
as described in Subsection~\ref{subsec:Jump_Diff}
using the parameter values in the captions of 
Tables~\ref{t:vg_ivol} and \ref{t:SubCEV_ivol}.
All computations in this subsection are performed using
the implementation 
in~\cite{LM_Code}
of our algorithm.
%In Section~\ref{sec:third_moments}

\begin{figure}[ht]
\begin{center}
\subfloat[\footnotesize{The probability distribution function for the spot
                 price 
		 $X_T$,
		 with the maturity
		 $T$
		 equal to 0.5, 1 and 2 years, 
		 where
	         $X$
		 is the Markov 
		 chain  used to approximate the variance gamma process.
		 For a precise description of the process
		 $X$
		 see Subsection~\ref{subsec:MCapp}.
		 All relevant parameter values
		 are given in the caption of 
		 Table~\ref{t:vg_ivol}.}]{\includegraphics[width=7.5cm]{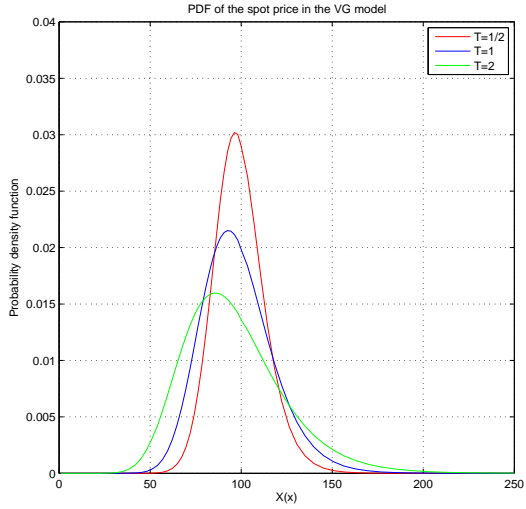} \label{fig:vg_pdf}}
\hspace{4mm}\subfloat[\footnotesize{The empirical probability distribution of the realized
    variance 
    $[\log(S)]_T$
    in the VG model
    $S$,
    based on the Monte Carlo simulation described in Subsection~\ref{subsec:VolDerJumps},
    and the distribution of the random variable 
    $I_T$
    for 
    $T\in\{0.5,1,2\}$
    matching
    $k\in\{2,3\}$
    instantaneous moments.
    For details on %the definition of
    $I_T$
    see Sections~\ref{sec:k_moments} and~\ref{sec:third_moments}.
    Note that the computational time required to obtain the law of
    $I_T$
    is independent of 
    $T$
    and that the quality of the approximation is greater for
    $k=3$
    (see also Table~\ref{t:vg_varpayoff}).}]{\includegraphics[width=8.3cm]{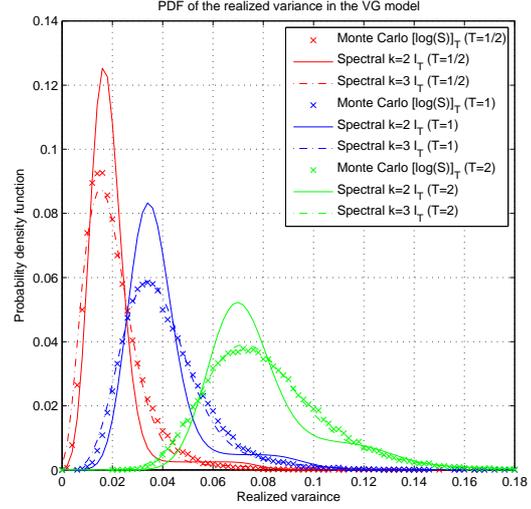}\label{fig:vg_var_pdf}}\\
\subfloat[\footnotesize{The function $x\mapsto\frac{4M_1(x)M_3(x)}{M_2(x)^2}$, for 
	$x\in E$ such that 
        $20\leq x\leq 250$, in the variance gamma model.
	This function appears 
	in 
	condition~\eqref{eq:max_Tx}
	of
	Subsection~\ref{subsec:JumpAlG}.
	The parameters of the chain 
	$X$
	are given in the caption of Table~\ref{t:vg_ivol}.}]{\includegraphics[width=7.8cm]{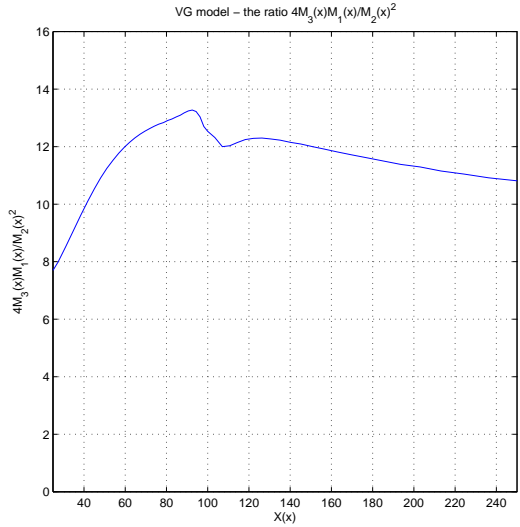}\label{fig:ratios}}
\hspace{4mm}\subfloat[\footnotesize{The functions
$x\mapsto\underline{\alpha}(x)$
and
$x\mapsto\overline{\alpha}(x)$,
for 
$x\in E$ 
such that 
$20\leq x\leq 250$, 
are the zeros of the quadratic
in condition~\eqref{eq:ieq_lambda_n}
in the variance gamma model.
As summarised in Table~\ref{t:conditions},
in order to ensure that
the intensity
$\lambda_n(x)$
is positive,
we must choose the value of the constant
$\alpha$
to lie 
between the two curves for
all
$x$
in the above range
(see also Subsection~\ref{subsec:VolDerJumps}).}]{\includegraphics[width=7.8cm]{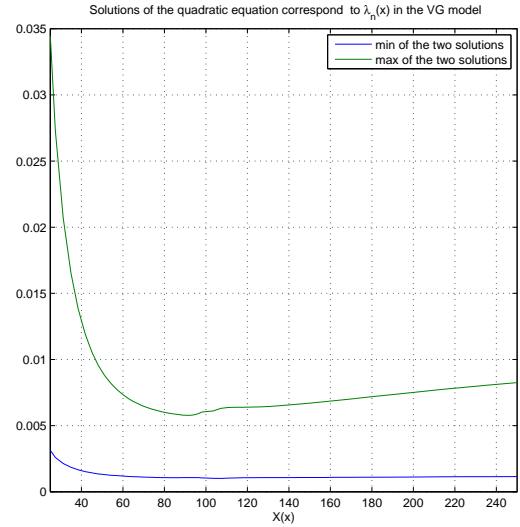}\label{fig:vg_solution_quad_eqn_lambda_n}}
\end{center}
\caption{Variance gamma model}
\end{figure}

Recall from Subsection~\ref{subsec:JumpAlG}
that in order to define the process
$I$
we need to set values for the integers
$1<n<m$
and the spacing
$\alpha$
so that the intensity
functions
$\lambda_1,\lambda_n,\lambda_m$
are positive
(Table~\ref{t:conditions}
states explicit necessary and sufficient conditions 
for this to hold).
Note that the inequality in~\eqref{eq:max_Tx}
is necessary if 
$\lambda_n$
is to be positive.
Figure~\ref{fig:ratios} 
contians the graph of the function
$x\mapsto 4M_1(x)M_3(x)/M_2(x)^2$
for 
$x\in E$
such that 
$20\leq x\leq 250$
in the case of the variance gamma model.
If we choose 
$$n:=5\quad
\text{and}\quad
m:=30,$$
then
the left-hand side of the inequality
in~\eqref{eq:max_Tx}
equals
$13.48$,
which is an upper bound for the ratio 
in Figure~\ref{fig:ratios}.
If
$S$
equals the subordinated CEV process, 
the graph of the function
$x\mapsto 4M_1(x)M_3(x)/M_2(x)^2$
takes a similar form and
the same choice
of 
$n,m$
as above 
satisfies the  inequality
in~\eqref{eq:max_Tx}.

The distance
$\alpha$
between the consecutive points in 
the state-space of the process
$I$
has to be chosen so that
the inequality 
$\underline{\alpha}(x) < \alpha < \overline{\alpha}(x)$
is satisfied for all
$x\in E$
(see Table~\ref{t:conditions}).
Figure~\ref{fig:vg_solution_quad_eqn_lambda_n}
contains the graphs of the functions 
$\underline{\alpha},\overline{\alpha}$
over the state-space of 
$X$
in the range
$20\leq x\leq 250$
for the variance gamma model.
The corresponding graphs 
in case of the subordianted CEV process
are very similar and are not reported. 
It follows that by choosing 
$$\alpha:=0.002$$
we can ensure that all the conditions in the third row of
Table~\ref{t:conditions}
are met,
both in the variance gamma and the subordinated CEV model,
for 
$x\in E$
such that 
$20\leq x\leq 250$.
%the states of the chain 
%$X$
%between
%$20$
%and
%$250$
It should be noted that it is impossible to find a single value of 
$\alpha$
that lies between the zeros 
$\underline{\alpha}(x)$
and
$\overline{\alpha}(x)$
for all 
$x\in E$
for our specific choice of the chain
$X$ and its state-space.
However not matching the instantaneous conditional
moments of 
$I_T$
and
$Q^{L,U}_T(X)$
outside of the interval
$[20,250]$
is 
in practice of little consequence because 
the probability that the chain
$X$
gets into this region 
%outside the interval
%$[20,250]$
(recall that the current spot level
is assumed to be 100)
before the maturity 
$T=2$
is negligible
(see Figure~\ref{fig:vg_pdf}
for the distribution of 
$X$
in the case of variance gamma model).

\begin{figure}[ht]
\begin{center}
\subfloat[\footnotesize{The distribution of the realized variance in the subordinated CEV model.}]{\includegraphics[width=9cm]{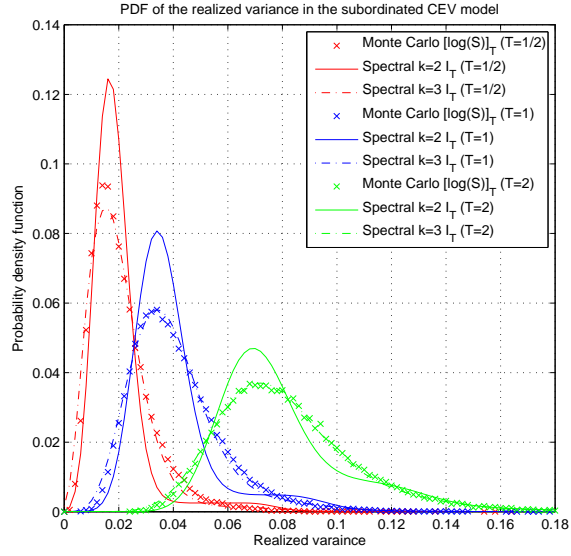}\label{fig:cevvg_var_pdf}}\hspace{1mm}
\subfloat[\footnotesize{The distribution of the corridor-realized variance in the subordinated CEV model.}]{\includegraphics[width=9cm]{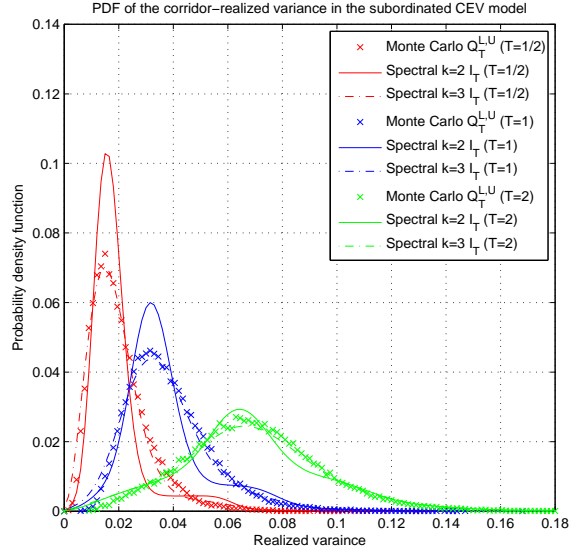}\label{fig:cevvg_corridor_var_pdf}}
\caption{\footnotesize{
Figure~\ref{fig:cevvg_var_pdf} (resp.~\ref{fig:cevvg_corridor_var_pdf})
contains
the empirical probability distribution of the realized
variance 
$[\log(S)]_T$
(resp. corridor-realized variance
$Q^{L,U}_T(S)$,
where
$L=70$
and
$U=130$)
in the subordinated CEV model
$S$,
based on the Monte Carlo simulation described in 
the caption of Table~\ref{t:SubCEV_ivol}
(see also Subsection~\ref{subsec:VolDerJumps}).
The distribution of the random variable 
$I_T$
for the maturity 
$T\in\{0.5,1,2\}$
matching
$k\in\{2,3\}$
instantaneous moments
is also plotted in both cases.
For details on %the definition of
$I_T$
see Sections~\ref{sec:k_moments} and~\ref{sec:third_moments}.
The computational time required to obtain the law of
$I_T$
is independent of 
$T$
and the quality of the approximation improves drastically for
$k=3$
(see also Tables~\ref{t:SubCEV_varpayoff}
and~\ref{t:SubCEV_corridor_varpayoff}).}}
\end{center}
\end{figure}

Once the parameters 
$n,m$
and
$\alpha$
have been determined, we use the explicit expressions for 
$\lambda_1,\lambda_n,\lambda_m$
on page~\pageref{SystemSol}
to define the state dependent intensities of the process
$I$
for the states 
$x\in E$
that satisfy
$20\leq x\leq 250$.
Outside of this region we choose the functions
$\lambda_1,\lambda_n,\lambda_m:E\to\RR_+$
to be constant.
The choice of parameter
$C=65$
is, like in the previous subsection, determined by the longest maturity
we are interested in (in our case this is
$T=2$).
The laws of the realized variance 
$[\log(S)]_T$,
for 
$T\in\{0.5,1,2\}$,
in the variance gamma
%\footnote{The implementation of the algorithm in Matlab
%in the case of the variance gamma model can be found at the 
%\href{http://www.ma.ic.ac.uk/~amijatov/Abstracts/eml.html}{\texttt{http://www.ma.ic.ac.uk/\~{}amijatov/Abstracts/eml.html}}}
%\texttt{http://www.ma.ic.ac.uk/$\sim$amijatov/}}
and the subordinated CEV model
based on the approximation
$I_T$
are given in Figures~\ref{fig:vg_var_pdf}
and~\ref{fig:cevvg_var_pdf}
respectively.
The prices of various payoffs on the realized variance in these two models 
are given in Tables~\ref{t:vg_varpayoff} and~\ref{t:SubCEV_varpayoff}.

\begin{table}[ht]
\begin{center}
\scalebox{0.8}{
\begin{tabular}{|c|c|ccc||ccc|ccc|}
\hline 
\textbf{VG}  &  $k$ moments & & Spectral: $I_T$ &  &  & MC: $[\log(S)]_T$ &  \\
\hline 
$\mathrm{derivative} \backslash T $ & & $0.5$ & $1$ & $2$ & $0.5$ & $1$ & $2$ \\
\hline
%%$\sqrt{{1\over T}\EE_0[\Sigma_T]}$ 
%var swap              & 1 &20.01\% & 20.01\% & 20.01\% & 20.00\% & 20.02\% & 20.02\%\\
% $\sqrt{\EE[\Sigma_T/T]}$              & 2 &20.01\% & 20.01\% & 20.01\% & (0.072\%) & (0.072\%) & (0.072\%) \\
%              & 3 &20.01\% & 20.01\% & 20.01\% &  &  &  \\
%\hline
%%$\EE_0\left[\sqrt{{1\over T}\Sigma_T}\right]$   
%vol swap              & 1 &19.75\% & 19.88\% & 19.95\% & 19.25\% & 19.62\% & 19.81\%\\
%$\EE\left[\sqrt{\Sigma_T/T}\right]$               & 2 &19.39\% & 19.66\% & 19.84\% & (0.024\%) & (0.018\%) & (0.013\%) \\
%              & 3 &19.26\% & 19.61\% & 19.80\% &  &  &  \\
%\hline
%call option& 1 &1.41\% & 1.36\% & 1.34\% & 1.65\% & 1.53\% & 1.47\%\\
% $\theta=80\%$              & 2 &1.56\% & 1.48\% & 1.45\% & (0.010\%) & (0.007\%) & (0.005\%) \\
%              & 3 &1.65\% & 1.53\% & 1.47\% &  &  &  \\
%\hline
%call option & 1 &0.50\% & 0.36\% & 0.25\% & 0.86\% & 0.63\% & 0.46\%\\
% $\theta=100\%$               & 2 &0.72\% & 0.57\% & 0.44\% & (0.008\%) & (0.005\%) & (0.004\%) \\
%               & 3 &0.83\% & 0.62\% & 0.46\% &  &  &  \\
%\hline
%call option & 1 &0.10\% & 0.05\% & 0.00\% & 0.38\% & 0.18\% & 0.07\%\\
%$\theta=120\%$               & 2 &0.33\% & 0.16\% & 0.06\% & (0.006\%) & (0.003\%) & (0.002\%) \\
%               & 3 &0.36\% & 0.19\% & 0.08\% &  &  &  \\
var swap              & 1 &	20.01\% &	20.01\% &	20.02\% &	20.01\% &	20.01\% &	20.01\%\\
$\sqrt{\EE[\Sigma_T/T]}$             
& 2 &	20.01\% &	20.01\% &	20.02\% &	(0.051\%) & (0.051\%) &	(0.051\%)\\
&  3 &	20.01\% &	20.01\% &	20.02\% &			 & &\\		
\hline
vol swap              & 1 &	19.74\% &	19.88\% &	19.96\% &	19.28\% &	19.62\% &	19.81\%\\
$\EE\left[\sqrt{\Sigma_T/T}\right]$              
& 2 &	19.40\% &	19.67\% &	19.83\% &	(0.017\%) & (0.012\%) &	(0.009\%)\\
&  3 &	19.25\% &	19.62\% &	19.81\% &		 & &\\		
\hline
call option & 1 &	1.51\% &	1.46\% &	1.44\% &	1.65\% &	1.52\% &	1.46\%\\
$\theta=80\%$             
& 2 &	1.56\% &	1.48\% &	1.45\% &	(0.007\%) & (0.005\%) &	(0.004\%)\\
&  3 &	1.66\% &	1.53\% &	1.47\% &			 & &\\		
\hline
call option & 1 &	0.50\% &	0.36\% &	0.25\% &	0.85\% & 0.63\%	& 0.45\%\\
$\theta=100\%$              
& 2 &	0.71\% &	0.56\% &	0.44\% &	(0.005\%) & (0.004\%) &	(0.003\%)\\
&  3 & 	0.83\% & 	0.61\% & 	0.45\% & 		 & &\\			
\hline
call option & 1 &	0.06\% &	0.01\% &	0.00\% &	0.37\% & 0.18\% & 0.07\%\\
$\theta=120\%$              
& 2 &	0.35\% &	0.22\% &	0.09\% &	(0.004\%) & (0.002\%) &	(0.001\%) \\
&  3 &	0.35\% &	0.18\% &	0.07\% &	 & &\\		
\hline
\hline
Time & & &  4s & &62s &120s &230s\\
\hline
\end{tabular}
}
\end{center}
\vspace{2mm}
\caption{\footnotesize{The prices of volatility derivatives in the variance gamma
model
$S$.  
The parameter values for the process 
$S$
and the chain
$X$
are given in the caption of Table~\ref{t:vg_ivol}.
The parameters for the process
$I$ 
are 
$\alpha=0.002$, $C=65$ 
for 
$k=1,2,3$.
We choose
$n=30$
when
$k=2$
and
$n=5, m=30$
when
$k=3$.
The variable
$\Sigma_T$
and the payoffs are as in Table~\ref{t:cev_varpayoff}.
%in the table
%denotes either 
%$I_T$
%or
%$[\log(S)]_T$
%and the call option price is
%$\EE\left[(\Sigma_T/T-(\theta K_0)^2)^+ \right]$,
%for 
%$\theta\in\{80\%, 100\%, 120\%\}$, 
%where
%$K_0:= \sqrt{\EE[\Sigma_T/T]}$.
The algorithm  
in~\cite{GlassMC}, page 144,
is used to generate
$10^5$
paths
of the VG process 
$S$
and the sum in~\eqref{eq:QuadVarS} is used
to obtain the empirical distribution of
$[\log(S)]_T$
(see Figure~\ref{fig:vg_var_pdf})
and to 
evaluate the contingent claims 
in this table. 
The numbers in brackets are the standard errors in 
the Monte Carlo simulation.
Note that the computational time for the pricing of volatility 
derivatives using the process
$I$
is independent of the maturity
$T$.
All computations are performed on the same hardware
as in Table~\ref{t:cev_ivol}
(see~\cite{LM_Code}
for the source code in Matlab).}}
\label{t:vg_varpayoff}
\end{table}

%In order to have a fair comparison between our methods with $k=2$ and $k=3$, we set the grid spacing $\alpha$, number of grid points $2C+1$
%and the maximum jump size multiple to be the same. We can see the prices computed by matching up to the third moment is much
%closer to the prices from the Monte Carlo method than the method with $k=2$. In Figure (\ref{fig:vg_var_pdf})
%we can also see that the pdfs for the realized variance
%obtained from the moment method with $k=3$ are much closer to the pdfs generated
%by the Monte Carlo method than the one computed by the second moment method ($k=2$).

\begin{table}[ht]
\begin{center}
\scalebox{0.8}{
\begin{tabular}{|c|c|ccc||ccc|ccc|}
\hline 
\textbf{CEV + jumps}  &  $k$ moments & & Spectral: $I_T$ &  &  & MC: $[\log(S)]_T$ &  \\
\hline 
$\mathrm{derivative} \backslash T $ & & $0.5$ & $1$ & $2$ & $0.5$ & $1$ & $2$ \\
\hline
var swap              
& 1 &	20.00\% &       20.03\% &	20.07\%	&        20.01\% &	20.03\% &	20.08\% \\
$\sqrt{\EE[\Sigma_T/T]}$             
& 2 &	20.00\% &	20.03\% &	20.07\% &	(0.051\%) &	(0.051\%) &	(0.051\%)\\
& 3 &	20.00\% &	20.03\% &	20.09\% &	 &		 &\\
\hline
vol swap              
& 1 &	19.73\% &	19.89\% &	19.98\% &	19.27\% &	19.63\% &	19.84\%\\
$\EE\left[\sqrt{\Sigma_T/T}\right]$              
& 2 &	19.39\% &	19.67\% &	19.85\% &	(0.017\%) &	(0.018\%) &	(0.010\%)\\
& 3 &	19.24\% &	19.62\% &	19.85\% &	 &		 &\\
\hline
call option 
& 1 &	1.51\% &	1.46\% &	1.45\% &	1.65\% &	1.53\% &	1.48\%\\
$\theta=80\%$             
& 2 &	1.56\% &	1.49\% &	1.46\% &	(0.007\%) &	(0.005\%) &	(0.004\%)\\
& 3 &	1.66\% &	1.54\% &	1.49\% &	 &		 &\\
\hline
call option 
& 1 &	0.51\% &	0.37\% &	0.30\% &	0.86\% &	0.64\% &	0.49\%\\
$\theta=100\%$              
& 2 &	0.71\% &	0.57\% &	0.47\% &	(0.006\%) &	(0.004\%) &	(0.003\%)\\
& 3 &	0.84\% &	0.63\% &	0.49\% &	 &		 &\\
\hline
call option 
& 1 &	0.06\% &	0.02\% &	0.01\% &	0.37\% &	0.19\% &	0.09\%\\
$\theta=120\%$              
& 2 &	0.36\% &	0.23\% &	0.11\% &	(0.004\%) &	(0.002\%) &	(0.001\%)\\
& 3 &	0.35\% &	0.19\% &	0.09\% &           &               & \\			
\hline
\hline
Time & & &  4s & &100s &200s &400s\\
\hline
\end{tabular}
}
\end{center}
\vspace{2mm}
\caption{\footnotesize{The prices of volatility derivatives in the subordinated CEV 
model
$S$.  
The parameter values for the process 
$S$
and the chain
$X$
are given in the caption of Table~\ref{t:SubCEV_ivol}.
The parameters for the process
$I$,
%are 
%given in Table~\ref{t:vg_varpayoff}.
%$\alpha=0.002$, $C=65$ 
%for 
%$k=1,2,3$.
%We choose
%$n=30$
%when
%$k=2$
%and
%$n=5, m=30$
%when
%$k=3$.
the random variable
$\Sigma_T$
and the payoffs of the volatility 
derivatives are as in Table~\ref{t:vg_varpayoff}.
%in the table
%denotes either 
%$I_T$
%or
%$[\log(S)]_T$
%and the call option price is
%$\EE\left[(\Sigma_T/T-(\theta K_0)^2)^+ \right]$,
%for 
%$\theta\in\{80\%, 100\%, 120\%\}$, 
%where
%$K_0:= \sqrt{\EE[\Sigma_T/T]}$.
The algorithm described in the caption of Table~\ref{t:SubCEV_ivol}
%in~\cite{GlassMC}, page 144,
is used to generate
$10^5$
paths
of the process 
$S$
and the sum in~\eqref{eq:QuadVarS} is used
to obtain the empirical distribution of
$[\log(S)]_T$
(see Figure~\ref{fig:cevvg_var_pdf})
and to 
evaluate the contingent claims 
in this table. 
The numbers in brackets are the standard errors in 
the Monte Carlo simulation.
Note that the computational time for the pricing of volatility 
derivatives using the process
$I$
is independent of the maturity
$T$.
All computations are performed on the same hardware
as in Table~\ref{t:cev_ivol}
(the code in~\cite{LM_Code}
can easily be adapted to this model).}}
\label{t:SubCEV_varpayoff}
\end{table}

Observe that the time required to compute the distribution of
$I_T$
in the case of the continuous process
$S$
(see Table~\ref{t:cev_varpayoff})
is larger than the time required to perform the equivalent task 
for the process with jumps
(see Tables~\ref{t:vg_varpayoff}
and~\ref{t:SubCEV_varpayoff}).
From the point of view of the algorithm this difference 
arises 
because
in the continuous case 
we have to use more points in the state-space of the process 
$I$
since
condition~\eqref{eq:FindAlphaDiff}
forces the choice of the smaller spacing 
$\alpha$.
In other words the 
quotient
$x\mapsto M_2(x)/M_1(x)$
takes 
much smaller values if there are no jumps
in the model 
$S$
than if there are.
It is intuitively clear from definition~\eqref{eq:CondMomentSigma}
that this ratio for the variance gamma (or the subordinated 
CEV) has a larger lower bound than 
the function in 
Figure~\ref{fig:cev_VQ_ratio},
because in the the diffusion case
the generator matrix is tridiagonal.
%Therefore including jumps in the model makes our algorithm faster.

\begin{table}[ht]
\begin{center}
\scalebox{0.8}{
\begin{tabular}{|c|c|ccc||ccc|ccc|}
\hline 
\textbf{CEV + jumps}  &  $k$ moments & & Spectral: $I_T$ &  &  & MC: $Q^{L,U}_T(S)$ &  \\
\hline 
$\mathrm{derivative} \backslash T $ & & $0.5$ & $1$ & $2$ & $0.5$ & $1$ & $2$ \\
\hline
corr-var swap              
& 1 &	19.81\% &       19.40\% &	18.50\%	&        19.81\% &	19.41\% &	18.50\% \\
$\sqrt{\EE[\Sigma_T/T]}$             
& 2 &	19.81\% &	19.40\% &	18.49\% &	(0.051\%) &	(0.050\%) &	(0.048\%)\\
& 3 &	19.81\% &	19.40\% &	18.50\% &	 &		 &\\
\hline
corr-vol swap              
& 1 &	19.59\% &	19.22\% &	18.25\% &	19.12\% &	19.03\% &	18.19\%\\
$\EE\left[\sqrt{\Sigma_T/T}\right]$              
& 2 &	19.18\% &	18.97\% &	18.08\% &	(0.016\%) &	(0.012\%) &	(0.005\%)\\
& 3 &	19.06\% &	18.93\% &	18.08\% &	 &		 &\\
%\hline
%call option 
%& 1 &	1.51\% &	1.46\% &	1.45\% &	1.64\% &	1.54\% &	1.49\%\\
%$\theta=80\%$             
%%& 2 &	1.56\% &	1.49\% &	1.46\% &	(0.010\%) &	(0.007\%) &	(0.006\%)\\
%& 3 &	1.67\% &	1.54\% &	1.48\% &	 &		 &\\
%\hline
%call option 
%& 1 &	0.51\% &	0.37\% &	0.30\% &	0.85\% &	0.64\% &	0.50\%\\
%$\theta=100\%$              
%%& 2 &	0.72\% &	0.58\% &	0.47\% &	(0.008\%) &	(0.005\%) &	(0.004\%)\\
%& 3 &	0.85\% &	0.63\% &	0.48\% &	 &		 &\\
%\hline
%%call option 
%& 1 &	0.06\% &	0.02\% &	0.01\% &	0.37\% &	0.19\% &	0.09\%\\
%$\theta=120\%$              
%& 2 &	0.37\% &	0.24\% &	0.11\% &	(0.006\%) &	(0.003\%) &	(0.002\%)\\
%& 3 &	0.36\% &	0.19\% &	0.09\% &           &               & \\			
\hline
\hline
Time & & &  4s & &100s &200s &400s\\
\hline
\end{tabular}
}
\end{center}
\vspace{2mm}
\caption{\footnotesize{Contingent claims on corridor-realized variance 
in the subordinated CEV 
model
$S$.  
The corridor is defined by
$L=70$
and
$U=130$.
All parameter values 
are as in Table~\ref{t:SubCEV_varpayoff}.
The empirical distribution
of
$Q^{L,U}_T(S)$
and the law of
$I_T$
for 
$T\in\{0.5,1,2\}$
are given in Figure~\ref{fig:cevvg_corridor_var_pdf}.
The Monte Carlo algorithm is as described in 
Table~\ref{t:SubCEV_ivol}
and the numbers in brackets are the standard errors in 
the simulation.}}
\label{t:SubCEV_corridor_varpayoff}
\end{table}

Finally we apply our algorithm to computing the law of the 
corridor-realized variance
$Q_T^{L,U}(S)$,
where
$S$
is the subordinated CEV process
and the corridor is given by
$L=70$
and
$U=130$.
It is clear from 
Figure~\ref{fig:cevvg_corridor_var_pdf}
and the price of the 
square root payoff  in
Table~\ref{t:SubCEV_corridor_varpayoff}
that the process 
$I$
defined by
matching 
$k=3$
instantaneous moments of
$Q^{L,U}_T(S)$
approximates best the entire distribution
of the corridor-realized variance.
However, if one is interested only in the value
of the corridor variance swap (i.e. a derivative with 
a payoff that is linear in
$Q^{L,U}_T(S)$),
Table~\ref{t:SubCEV_corridor_varpayoff}
shows that 
it suffices to take
$k=1$.

\section{Conclusion}
\label{sec:conclusion}
We proposed an algorithm for pricing and hedging
volatility derivatives and derivatives on the 
corridor-realized variance in markets driven by
Markov processes of dimension one.
The scheme is based on an order 
$k$
approximation of the corridor-realized variance process
by a continuous-time Markov chain.
We proved the weak convergence of our scheme
as
$k$
tends to infinity
and demonstrated with numerical examples that in practice 
it is sufficient to use 
%it behaves well for 
$k=2$
if the underlying Markov process is continuous
and 
$k=3$
if the market model has jumps.

There are two natural open questions related to this algorithm.
First,
%from the theoretical point of view 
it would be interesting to understand 
the precise rate of convergence in Theorem~\ref{thm:Conv}
both from the theoretical point of view and that of 
applications. 
The second question is numerical in nature.
As mentioned in
the introduction, the algorithm described in this paper can be adapted to
the case when the process 
$S$
is a component of a two dimensional Markov process. The implementation
of the algorithm in this case is hampered by the dimension of the generator
of the approximating Markov chain, which would in this case be approximately 
$2000$
(as opposed to 
$70$,
as
in the examples of Section~\ref{sec:numerical}).
It would be interesting to understand the precise structure of
this large generator matrix and perhaps exploit it to obtain an efficient
algorithm for pricing volatility derivatives in the presence
of stochastic volatility.

\appendix

\section{Partial-circulant matrices}
\label{sec:Bloc}
A matrix
$C\in\RR^{n\times n}$
is \textit{circulant}
if 
there exists a vector
$c\in\RR^n$
such that
%it is
%of the form
%\[
%C=
%\begin{pmatrix}
%c_0     & c_1     & c_2    & \cdots & c_{n-1} \\
%c_{n-1} & c_0     & c_1    & \cdots & c_{n-2} \\
%c_{n-2} & c_{n-1} & c_0    & \ddots & \vdots \\
%\vdots  & \vdots  & \ddots & \ddots & c_1 \\
%c_1     & c_2     & \cdots & c_{n-1}& c_0 
%\end{pmatrix},
%\]
%where each row is a cyclic permutation of the row above.
%In other words
%an entry 
%$C_{ij}$
%of the matrix
%$C$
%can be expressed as
$C_{ij}=c_{(i-j)\!\!\!\mod n}$
for all
$i,j\in\{1,\ldots,n\}.$
The matrix
$C$
can always be diagonalised
analytically, when viewed as a linear operator
on the complex vector space
$\CC^n$,
as follows.
For any 
$r\in\{0,\ldots,n-1\}$
we have an eigenvalue
$\lambda_r$
and
a corresponding eigenvector
$y^{(r)}$
(i.e.
the equation
$Cy^{(r)}=\lambda_r y^{(r)}$
holds for all 
$r$
and the family of vectors
$y^{(r)}$,
$r\in\{0,\ldots,n-1\}$,
spans the whole of
$\CC^n$)
of the form
\begin{eqnarray*}
\lambda_r  =  \sum_{k=0}^{n-1}c_k e^{-i\frac{2\pi}{n}rk}\quad\text{and}\quad
y^{(r)}_j  =  \frac{1}{\sqrt{n}}e^{-i\frac{2\pi}{n}rj}
\>\>\>\>\mathrm{for}\>\>\>\>j\in\{0,\ldots,n-1\}.
\end{eqnarray*}
It is interesting to note that the eigenvectors
$y^{(r)}$,
$r\in\{0,\ldots,n-1\}$,
are independent of the circulant matrix 
$C$.
For the proof of these statements see Appendix~A
in~\cite{ALM}.

Let
$A$
be a linear operator represented by a matrix
in
$\RR^{m\times m}$
and let
$B^{(k)}$,
for
$k=0,\ldots,m-1$,
be a family of
$n$-dimensional matrices with
the following property: there exists an invertible matrix
$U\in\CC^{n\times n}$
such that
$$U^{-1}B^{(k)}U=\Lambda^{(k)},\>\>\>\>\mathrm{for}\>\>\mathrm{all}\>\>\>\>
k\in\{0,\ldots,m-1\},$$
where
$\Lambda^{(k)}$
is a diagonal matrix in
$\CC^{n\times n}$.
In other words this condition stipulates
that the family of matrices
$B^{(k)}$
can be
simultaneously diagonalized by the transformation
$U$.
Therefore the columns of matrix
$U$
are eigenvectors of
$B^{(k)}$
for all
$k$
between
0
and
$m-1$.

Let us now define a large linear operator
$\widetilde{A}$,
acting on a vector space of dimension
$mn$,
in the following way. Clearly the matrix
$\widetilde{A}$
can be decomposed naturally into
$m^2$
blocks of size
$n\times n$.
Let
$\widetilde{A}_{i,j}$
denote an
$n\times n$
matrix which represents
the block in
the
$i$-th row and
$j$-th column
of this decomposition.
We now define the operator
$\widetilde{A}$
as
\begin{eqnarray}
\label{eq:diag}
\widetilde{A}_{ii} & := & B^{(i)}+A_{ii}\II_{\RR^n}\>\>\>\>\mathrm{and}\\
\label{eq:diag1}
\widetilde{A}_{ij} & := & A_{ij}\II_{\RR^n},\>\>\>\>\mathrm{for}\>\>\mathrm{all}
\>\>\>\>i,j\in\{1,\ldots,m\}\>\>\>\>\mathrm{such}\>\>\mathrm{that}\>\>\>\>
i\neq j.
\end{eqnarray}
The real numbers
$A_{ij}$
are the entries of matrix
$A$
and
$\II_{\RR^n}$
is the identity operator
on
$\RR^n$.
We may now state our main definition.

\begin{defin}
A matrix is termed
\textit{partial-circulant}
if it admits a structural decomposition
as in~(\ref{eq:diag}) and~(\ref{eq:diag1})
for any matrix
$A\in\RR^{m\times m}$
and a family of
$n$-dimensional
circulant
matrices
$B^{(k)}$,
for
$k=0,\ldots,m-1$.
\end{defin}

\noindent For the spectral properties of partial-circulant matrices
see Appendix~A
in~\cite{ALM}.

\section{Non-uniform state-space of the Markov chain $X$}
\label{sec:NU_State_Space}
%We are going to present some numerical results to demonstrate the accuracy
%of our method. First of all, we need to define an algorithm to generate a
%non-uniform grid. 
The task here is to construct a
non-uniform state-space for the Markov chain
$X$, 
which was
used in Section~\ref{sec:numerical}
to approximate the Markov process 
$S$.
%A widely used method for generating a non-uniform state-space consists of 
%applying a coordinate transformation to a uniform state-space, which is what we will
%do.
Recall that the state-space is 
a set of non-negative real numbers
$E=\{x_0,x_1,\ldots,x_{N-1}\}$
for some even integer
$N\in2\NN$.
Recall that 
the elements of the set
$E$,
when viewed as a finite sequence,
are strictly increasing.
We first fix
three real numbers 
$l,s,u\in\RR$,
such that
$l<s<u$,
that specify the boundaries of the lattice 
$x_0=l$,
$x_{N-1}=u$
and the starting point of the chain
$x_{\lceil N/2\rceil} =s=S_0$
which coincides with the initial spot value in the model
$S$.
The function
$\lceil\cdot\rceil :\RR\to\ZZ$
returns the smallest integer which is larger or equal than the
argument.
We next choose strictly positive parameter values
$g_l,g_u$ 
which control the granularity of the spacings between 
$l$ and 
$s$ and between 
$s$ and 
$u$
respectively. 
In other words the larger 
$g_l$
(resp.
$g_u$)
is,
the more uniformly spaced the lattice is 
in the interval
$[l,s]$
(resp.
$[s,u]$).
%Let 
%$N$ be the number of
%grid points. The points 
%$X(0)=l$ 
%and 
%$X(N-1)=u$ 
%define the lower and upper boundary of the domain respectively. 
%The mid point 
%$X\left(\lceil N/2 \rceil\right)=s$ 
%corresponds to the initial spot value. 
The algorithm that constructs the lattice points 
is a slight modification of
the algorithm in~\cite{TR}, page 167,  
and can be described as follows.

\begin{enumerate}
\item Compute $c_1 = \mathrm{arcsinh}\left(\frac{l-s}{g_l} \right)$, 
           $c_2 = \mathrm{arcsinh}\left(\frac{u-s}{g_u} \right)$,  
	   $N_l = \lceil N/2\rceil$ and $N_u = N-(N_l+1)$.
%\item Generate a uniform grid $y\in[0,1]$ with $N_l$ number of points.
\item Define the lower part of the grid by the formula
      $x_k := s + g_l \mathrm{sinh}(c_1(1-k/N_l))$
      for 
      $k\in\{0,\ldots,N_l\}$.
      Note that $x_0=l, x_{N_l}=s$.
%\item Generate a uniform grid $y\in[0,1]$ with $N_u$ number of points.
\item Define the upper part of the grid using the formula
      $x_{N_l+k} := s + g_u \mathrm{sinh}(c_2 k/N_u)$
      for 
      $k\in\{0,\ldots,N_u\}$.
      Note that 
      $ x_{N-1}=u$.
%\item Define the non-uniform lattice by concatenating 
%      the lower and upper parts given in (3)~and~(4). 
\end{enumerate}

\bibliographystyle{plain}
\bibliography{cite}
%\nocite{*}

\end{document}